\documentclass[prc,aps,float,amsmath,amssymb,amsfonts,nofootinbib,superscriptaddress,twocolumn]{revtex4-1}
\usepackage{graphicx}
\usepackage[usenames,dvipsnames]{color}
\usepackage[mathscr]{eucal}

\newcommand{\N}{\mathcal{N}}
\renewcommand{\P}{\mathcal{P}}
\renewcommand{\arraystretch}{1.2}

\begin{document}
\title{Particle-number projected Bogoliubov coupled cluster theory. \\ Application to the pairing Hamiltonian}
\author{Y. Qiu}
\affiliation{Department of Chemistry, Rice University, Houston, TX 77005-1892}

\author{T. M. Henderson}
\affiliation{Department of Chemistry, Rice University, Houston, TX 77005-1892}
\affiliation{Department of Physics and Astronomy, Rice University, Houston, TX 77005-1892}

\author{T. Duguet}
\affiliation{IRFU, CEA, Universit\'e Paris - Saclay, F-91191 Gif-sur-Yvette, France}
\affiliation{KU Leuven, Instituut voor Kern- en Stralingsfysica, 3001 Leuven, Belgium}

\author{G. E. Scuseria}
\affiliation{Department of Chemistry, Rice University, Houston, TX 77005-1892}
\affiliation{Department of Physics and Astronomy, Rice University, Houston, TX 77005-1892}

\date{Updated \today}

\begin{abstract}
\begin{description}
\item[Background] While coupled cluster theory accurately models weakly correlated quantum systems, it often fails in the presence of strong correlations where the standard mean-field picture is qualitatively incorrect.  In many cases, these failures can be largely ameliorated by permitting the mean-field reference to break physical symmetries. Symmetry-broken coupled cluster, e.g. Bogoliubov coupled cluster, theory can indeed provide reasonably accurate energetic predictions, but the broken symmetry can compromise the quality of the resulting wave function and predictions of observables other than the energy.
\item[Purpose] Merging symmetry projection and coupled cluster theory is therefore an appealing way to describe strongly correlated systems. One indeed expects to inherit and further improve the energetic accuracy of broken-symmetry coupled cluster while retaining proper symmetries. 
\item[Methods] Independently, two different but related formalisms have been recently proposed to achieve this goal.  The two formalisms are contrasted in this manuscript, with results tested on the Richardson pairing Hamiltonian. While the present paper  focuses on the breaking and restoration of $U(1)$ global-gauge symmetry associated with particle number conservation, the symmetry-projected coupled cluster formalism is applicable to other symmetries such as rotational (i.e. spin) symmetry.
\item[Results] Both formalisms are based on the disentangled cluster representation of the symmetry-rotated coupled cluster wavefunction. However, they differ in the way that the disentangled clusters are solved. One approach sets up angle-dependent coupled cluster equations, while the other involves first-order ordinary differential equations. The latter approach yields energies and occupation probabilities significantly better than those of number-projected BCS and BCS coupled cluster and, when the disentangled clusters are truncated at low excitation levels, has a computational cost not too much larger than that of BCS coupled cluster.
\item[Conclusions] The high quality of results presented in this manuscript indicates that symmetry-projected coupled cluster is a promising method that can accurately describe both weakly and strongly correlated finite many-fermion systems.
\end{description}
\vspace{3pt}

\end{abstract}

\maketitle

\section{Introduction
\label{intro}}

The description of weakly correlated quantum systems typically starts with a mean-field approximation providing a qualitative but not quantitative characterization of the problem before incorporating many-body correlations via a method of choice.  Coupled cluster (CC) theory \cite{Coester1958,Cizek1966,Paldus1999,Bartlett2007,ShavittBartlett} is one of the most popular and powerful ways to do so, as it accurately accounts for these correlation effects at a reasonable computational cost that scales polynomially with system size.

Strongly correlated systems are much more challenging, but mean-field methods are surprisingly resilient. Indeed, the mean-field picture can, by breaking some or even all of the symmetries of the system, correctly describe at least some of the important qualitative physics in such a situation. For example, dealing with the pairing or reduced BCS Hamiltonian via the breaking of $U(1)$ global gauge symmetry associated with particle-number conservation, mean-field theory correctly occupies every single-particle level.  The symmetry-breaking mean-field state can then serve as a reference to expand the exact ground-state wave-function and, in many strongly correlated problems, deliver accurate energies at a reasonable price.

However, methods based on a symmetry-breaking mean-field reference state have important shortcomings. They do not account for the entanglement which is a hallmark of strong correlations. Indeed, quantum fluctuations do eventually lift the fictitious degeneracy associated with the broken symmetry given that the latter is in fact only {\it emergent} in finite systems~\cite{ui83a,yannouleas07a,Papenbrock:2013cra}. It is thus mandatory to restore good symmetry quantum numbers, which not only revises the energy (dramatically in certain situations) but also allows the proper handling of transition operators characterized by symmetry selection rules. The shortcomings are diminished but not resolved in high-orders broken-symmetry, e.g. coupled cluster, methods given that they can in fact only restore the broken symmetry in an all-order limit.

At the mean-field level, it has long been realized that the qualitative failures of breaking symmetries are largely resolved by symmetry projection \cite{Lowdin55c,pauncz1967,mayer1980,ring80a,Blaizot85,bender03b,Schmid2004,PHF}, at least in finite systems. Indeed, the action of the projector retains only that portion of the symmetry-breaking mean-field state that has the correct symmetry properties.  It seems natural to extend this idea to the case of symmetry-breaking coupled cluster theory.
While there have been attempts to merge a symmetry-projected mean-field reference with a symmetry-adapted cluster operator \cite{Piecuch1996,Qiu2017}, practical attempts to projectively restore the symmetries of a broken-symmetry coupled cluster wave function have only been made in the past few years.  
Duguet \cite{Duguet:2014jja,Duguet:2015yle} and Scuseria \cite{qiu17a} have proposed independent ways to achieve this, and Tsuchimochi and Ten-no have provided a third approach \cite{Tsuchimochi2017}, albeit so far at the linearized coupled cluster level.

The first goal of this paper is to discuss the formal relationship between the approaches introduced in Refs.~\cite{Duguet:2014jja,Duguet:2015yle} and in Ref.~\cite{qiu17a}.  While these two methods were originally formulated in different ways, it is presently shown that they follow closely related ideas and can be formulated within the same basic framework.  We seek to standardize the language and reveal the similarities between both methods, as well as to highlight their differences.

The second goal of the present paper is to test symmetry-projected coupled cluster theory on the reduced BCS or pairing Hamiltonian. In this context, the broken and restored symmetry is $U(1)$ global-gauge symmetry associated with the conservation of particle number. Indeed, the mean-field solutions generated in the strongly correlated regime of the pairing Hamiltonian are particle-number breaking BCS states. Although the results are restricted to the schematic pairing Hamiltonian, the formalism presented in this manuscript is directly applicable to the more general context of Bogoliubov coupled-cluster \cite{Henderson:2014vka,Signoracci:2014dia} such that the  applied method is coined as particle-number projected Bogoliubov coupled cluster (PBCC) theory.

The paper is organized as follows. In Sec.~\ref{Sec:many-body}, the main features of the PBCC formalism are introduced. Additional formal and technical details are supplied in set of appendices. While the presentation remains general in Sec.~\ref{Sec:many-body}, it is further specified to the pairing Hamiltonian in App.~\ref{PH}.  Numerical results obtained for the pairing Hamiltonian are discussed in Sec.~\ref{Sec:application} while conclusions are given in Sec.~\ref{Sec:Conclusions}.

\section{Many-body formalism}
\label{Sec:many-body}

In addition to laying down the formalism in view of applying it, the goal of the present section is to clarify the structural identity of the two versions of the PBCC formalism introduced independently in Refs.~\cite{Duguet:2014jja,Duguet:2015yle} and~\cite{qiu17a} on the one hand, and to compare the equations put forward in both cases to solve for the gauge-angle-dependent cluster amplitudes on the other hand.

\subsection{Bogoliubov coupled cluster}

Before coming to PBCC theory, it is first necessary to briefly outline the underlying Bogoliubov coupled cluster (BCC) formalism~\cite{Henderson:2014vka,Signoracci:2014dia}. Bogoliubov coupled cluster begins with the introduction of a Bogoliubov reference state
\begin{equation}
\label{e:bogvac}
| \Phi \rangle \equiv \mathcal{C} \displaystyle \prod_{k} \beta_{k} | 0 \rangle \, ,
\end{equation}
which is a vacuum for the set of quasiparticle operators $\beta$ and $\beta^\dagger$ obtained from particle ones $c$ and $c^\dagger$ via a linear Bogoliubov transformation~\cite{ring80a}
\begin{equation}
\begin{pmatrix} \beta \\ \beta^\dagger \end{pmatrix} = \mathcal{W} \, \begin{pmatrix} c \\ c^\dagger \end{pmatrix} = \begin{pmatrix} \mathcal{U}^\dagger & \mathcal{V}^\dagger \\ \mathcal{V}^\mathsf{T} & \mathcal{U}^\mathsf{T} \end{pmatrix} \, \begin{pmatrix} c \\ c^\dagger \end{pmatrix}
\end{equation}
where the unitary Bogoliubov transformation $\mathcal{W}$ is typically obtained by solving Hartree-Fock-Bogoliubov (HFB) equations but does not need to be. While $| \Phi \rangle$ reduces to a Hartree-Fock (HF) Slater determinant when $\mathcal{V}=0$, it is in general not an eigenstate of the particle number operator\footnote{The reader is advised not to be confused between the operator $A$ (math style) and its eigenvalue A (roman style) used throughout the paper.} $A$, i.e. it breaks $U(1)$ global gauge symmetry.

Many-body correlations are incorporated via the usual exponential ansatz of coupled cluster theory
\begin{equation}
|\Psi_\mathrm{BCC}\rangle \equiv e^U |\Phi\rangle = \prod_{n=1} e^{U_n} |\Phi\rangle,
\label{Eqn:PsiBCC}
\end{equation}
where $U_n$ is a $n$-tuple (i.e. $2n$ quasiparticle) connected excitation operator with respect to $| \Phi \rangle$
\begin{equation}
U_n \equiv \frac{1}{(2n)!} \!\!\sum_{k_1\ldots k_{2n}}\!\! U^{2n0}_{k_1\ldots k_{2n}} \beta^\dagger_{k_1} \ldots \beta^\dagger_{k_{2n}},
\label{clusteropntuple}
\end{equation}
such that $[U_p,U_q]=0$.  Defining the n-tuple excitation operator\footnote{The $n$-tuple (i.e. $2n$ quasiparticle) de-excitation operator is consistently defined as $\mathcal{B}_\mu \equiv \left(\mathcal{B}^\mu\right)^\dagger = \beta_{k_{2n}} \ldots \beta_{k_1}$.}
\begin{equation}
\mathcal{B}^\mu \equiv \mathcal{B}^{k_1 \ldots k_{2n}} = \beta_{k_1}^\dagger \ldots \beta_{k_{2n}}^\dagger \, ,
\end{equation}
allows one to introduce $n$-tuple excitations of the Bogoliubov vacuum through
\begin{equation}
| \Phi^{\mu} \rangle  \equiv \mathcal{B}^\mu | \Phi \rangle \, .
\end{equation}

With those definitions at hand, the insertion of the BCC wave function ansatz into the Schr\"odinger equation yields the BCC energy and amplitude equations
\begin{subequations}
\begin{align}
E^{\text{A}} &= \langle \Phi| H \, e^U |\Phi\rangle \, , \label{Eqn:BCCEn} \\
0 &= \langle \Phi^{\mu} | \left(H - E^{\text{A}}\right) \, e^U |\Phi\rangle \, , \label{Eqn:BCCAmp}
\end{align}
\label{Eqn:BCCAll}
\end{subequations}
where intermediate normalization $\langle \Phi|\Psi_\mathrm{BCC}\rangle = \langle \Phi| e^U |\Phi\rangle  =1$ was used in Eq.~\ref{Eqn:BCCEn}.

These equations are more usually formulated in terms of a similarity-transformed Hamiltonian, but these similarity-transformed equations are equivalent to those presented in the foregoing.  The amplitudes $U^{2n0}_\mu$ are obtained by solving Eq. \ref{Eqn:BCCAmp}, which constitutes a set of coupled nonlinear equations.

\subsection{Particle number projection}

If BCC Eq. \ref{Eqn:BCCAmp} is satisfied for all $\mu$, the BCC wave function is an exact eigenfunction of the Hamiltonian.  Provided that the Hamiltonian governing the A-body quantum system is a particle-number conserving operator such that $[A,H] = 0$, elementary considerations dictate that in the absence of degeneracies the eigenstates of the Hamiltonian $H$ carry good symmetry quantum number A, i.e. are particle number eigenfunctions.  The BCC wave function is thus an eigenfunction of both $H$ and $A$.

In practical calculations, however, the expansion of $U = \sum_n U_n$ must be truncated, and generally only a subset of the BCC amplitude equations can then be satisfied.  In this case, the BCC wave function is no longer a particle number eigenfunction.  The idea of PBCC is to remedy this by applying a particle number projection operator to the BCC wave function\footnote{When all BCC amplitude equations are satisfied, the number projection is obviously harmless and $|\Psi_\mathrm{PBCC}\rangle=\nobreak|\Psi_\mathrm{BCC}\rangle$.}
\begin{equation}
|\Psi_\mathrm{PBCC}\rangle \equiv P^{\text{A}} |\Psi_\mathrm{BCC}\rangle.
\label{PCCansatz}
\end{equation}
The projection operator on particle number $\text{A}$ reads as
\begin{align}
P^{\text{A}} &\equiv  \frac{1}{2\pi}\int_{0}^{2\pi} \!d\varphi \, e^{-i\text{A}\varphi}  \, R(\varphi) \, , \label{projAoperator}
\end{align}
where the operator $R(\varphi)\equiv e^{iA\varphi}$ forms the abelian compact Lie group $U(1)\equiv \{R(\varphi), \varphi \in  [0,2\pi]\}$ associated with the global rotation of the A-body fermion system in gauge space.

Let us note that it is helpful to use a variation after projection (VAP) scheme in which the Bogoliubov transformation $\mathcal{W}$ minimizes the projected mean-field energy rather than the straight mean-field one. Results presented in Sec.~\ref{Sec:Results} are all based on a Bogoliubov transformation obtained in this way.

\subsection{Ground-state energy}
\label{energy}

Inserting the PBCC wave-function ansatz into the Schr\"odinger equation and commuting the Hamiltonian with the projection operator leads to
\begin{align}
\langle \Phi |P^{\text{A}} H e^{U} | \Phi \rangle &=  E^{\text{A}} \, \langle \Phi |P^{\text{A}} e^{U} | \Phi \rangle.
\label{grandpoteigenvaleq} 
\end{align}
Expanding $P^{\text{A}}$ according to Eq.~\ref{projAoperator} and introducing off-diagonal, i.e. left-rotated, unexcited norm and Hamiltonian kernels
\begin{subequations}
\label{kernelsA}
\begin{align}
{\cal N}(\varphi)  &\equiv \langle \Phi(\varphi) | e^{U} | \Phi \rangle \, , \label{kernelsA1} \\
{\cal H}(\varphi) &\equiv \langle \Phi(\varphi) |  H e^{U} | \Phi \rangle  \, , \label{kernelsA2}
\end{align}
\end{subequations}
where the gauge-rotated Bogoliubov bra state is defined as $\langle \Phi(\varphi) |\equiv \langle \Phi |R(\varphi)$, the PBCC energy is written as
\begin{align}
E^{\text{A}} &=   \frac{\int_{0}^{2\pi} \!d\varphi \, e^{-i\text{A}\varphi}  \, {\cal H}(\varphi)}{\int_{0}^{2\pi} \!d\varphi \, e^{-i\text{A}\varphi}  \, {\cal N}(\varphi)}  \, . \label{projeigenequatkernelsA1} 
\end{align}

Expressing $\langle\Phi(\varphi) |$ via a Thouless transformation of $\langle\Phi |$ (see App.~\ref{SecThouless} for details)
\begin{equation}
\label{thoulessbetweenbothvacua2}
\langle \Phi(\varphi) | \equiv \langle \Phi(\varphi) | \Phi \rangle \langle \Phi | e^{Z(\varphi)} \, ,
\end{equation}
where the Thouless operator $Z(\varphi)$ is a pure de-excitation operator when acting to the right (and a pure excitation operator when acting to the left), the two kernels can be rewritten as
\begin{subequations}
\label{kernelsB}
\begin{align}
{\cal N}(\varphi)  &\equiv \langle\Phi(\varphi) | \Phi \rangle \langle \Phi | e^{U_{Z}(\varphi)} | \Phi \rangle \, , \label{kernelsB1} \\
{\cal H}(\varphi) &\equiv \langle\Phi(\varphi) | \Phi \rangle \langle \Phi |  H_{Z}(\varphi) e^{U_{Z}(\varphi)} | \Phi \rangle  \, , \label{kernelsB2} 
\end{align}
\end{subequations}
where use was made of the identity $e^{Z(\varphi)}| \Phi \rangle=| \Phi \rangle$. The operators entering Eq.~\ref{kernelsB} are defined through the similarity transformation
\begin{subequations}
\label{transformedop}
\begin{align}
H_{Z}(\varphi) &\equiv e^{Z(\varphi)} H e^{-Z(\varphi)} \, , \label{transformedH} \\
e^{U_{Z}(\varphi)} &\equiv e^{Z(\varphi)} e^{U} e^{-Z(\varphi)}  \, , \label{transformedexpU}
\end{align}
\end{subequations}
which is discussed in details in App.~\ref{Sectransforop}.

The angle-dependent state $\exp(U_Z(\varphi)) |\Phi\rangle$ can be written in coupled cluster form, as
\begin{equation}
\label{WDef}
e^{U_{Z}(\varphi)} | \Phi \rangle = e^{W(\varphi)} | \Phi \rangle.
\end{equation}
Here, the gauge-angle-dependent operator $W(\varphi)\equiv\sum_{n=0}W_n(\varphi)$ is made out of $n$-fold excitation operators and includes a normalization constant $W_0(\varphi)$.  They can be obtained by expanding $e^{U_{Z}(\varphi)}$, normal ordering each term with respect to $| \Phi \rangle$ and retaining the non-zero terms of $e^{U_{Z}(\varphi)} | \Phi \rangle$ before rewriting the resulting wave function in coupled cluster form.  The excitations $W_n(\varphi)$ are denoted as disentangled clusters, because this process above is equivalent to disentangling the algebra formed by excitations, de-excitations and quasiparticle number-conserving operators. Adopting the language of disentangled clusters allows one to rewrite Eq.~\ref{kernelsB} as 
\begin{subequations}
\label{kernelsC}
\begin{align}
{\cal N}(\varphi)  &= \langle\Phi(\varphi) | \Phi \rangle e^{W_0(\varphi)} \, , \label{kernelsC1} \\
{\cal H}(\varphi) &= \langle \Phi | H_{Z}(\varphi) e^{{\cal T}(\varphi)} | \Phi \rangle  {\cal N}(\varphi)\, , \label{kernelsC2} 
\end{align}
\end{subequations}
where ${\cal T}(\varphi)\equiv W(\varphi) - W_0(\varphi)$ and with
\begin{align}
W_n(\varphi) &\equiv \left[e^{U_{Z}(\varphi)}\right]^{2n0}_{C} \nonumber  \\
&= \sum_{l=0}^{\infty} \frac{1}{l!} \left[(U_{Z}(\varphi))^l\right]^{2n0}_{C} \nonumber \\
&\equiv \frac{1}{2n!} \!\!\sum_{k_1\ldots k_{2n}}\!\! W^{2n0}_{k_1\ldots k_{2n}}(\varphi) \beta^\dagger_{k_1} \ldots \beta^\dagger_{k_{2n}}  \, , \label{angledependentclusters}
\end{align}
where $C$ denotes the normal-ordered part of n-tuple excited contributions to $e^{U_{Z}(\varphi)}$ obtained by contracting strings of $U_k$ operators. It is to be noted that $W_n(\varphi)$ receives contributions from all $U_k$ and from all excitation ranks in the expansion of $e^{U_{Z}(\varphi)}$.  Morever, even if $U$ is truncated at some low order of excitation, $W$ will generally be non-zero for all excitation orders.  Practical calculations thus require that both the broken-symmetry cluster operator $U$ and the disentangled cluster operator $W(\varphi)$ be truncated, or at least that higher-order cluster operators be approximated in terms of lower-order ones.

While Eq.~\ref{angledependentclusters} provides a formal definition of the gauge-rotated cluster operators $W_n(\varphi)$, it is not advocated to access them in this way in the following, essentially because the summation over $l$ in the second line of Eq.~\ref{angledependentclusters} converges too slowly to be of use.

Equation~\ref{kernelsC} demonstrates the typical factorization property of off-diagonal kernels that leads to introducing the {\it correlated} part of ${\cal N}(\varphi)$ and the {\it connected} part of ${\cal H}(\varphi)$ via
\begin{subequations}
\label{reducedkernels}
\begin{align}
n(\varphi)  &\equiv \frac{{\cal N}(\varphi)}{\langle\Phi(\varphi) | \Phi \rangle} = e^{W_0(\varphi)} \, , \label{reducedkernels1} \\
h(\varphi) &\equiv \frac{{\cal H}(\varphi)}{{\cal N}(\varphi)} = \langle \Phi |  H_{Z}(\varphi) e^{{\cal T}(\varphi)} | \Phi \rangle_{C} \, , \label{reducedkernels2} 
\end{align}
\end{subequations}
where the index $C$ stipulates the connected character of the kernel. 

With all these developments at hand, the final form of the PBCC energy reads as
\begin{align}
E^{\text{A}} &= \frac{\int_{0}^{2\pi} \!d\varphi \, e^{-i\text{A}\varphi}  \, h(\varphi){\cal N}(\varphi)}{\int_{0}^{2\pi} \!d\varphi \, e^{-i\text{A}\varphi}  \, {\cal N}(\varphi)} \label{projeigenequatkernelsB1} \, .
\end{align}

\subsection{Discussion}
\label{discussion1}

Equations~\ref{kernelsC}, \ref{reducedkernels} and~\ref{projeigenequatkernelsB1} stipulating the way the energy is to be computed in PBCC theory appear identically in Ref.~\cite{Duguet:2014jja,Duguet:2015yle} and~\cite{qiu17a}. Both approaches make the same use of the similarity-transformed operator $H_Z(\varphi)$, once recognized that the Thouless operator $Z(\varphi)$ corresponds to both the operator $V_1$ introduced in Ref.~\cite{qiu17a} and to the operator $(R^{--}(\varphi))^{\dagger}$ introduced in Ref.~\cite{Duguet:2015yle}. Furthermore, $n(\varphi) = \exp(W_0(\varphi))$ and the disentangled clusters appear identically once recognized that ${\cal T}(\varphi)$~\cite{Duguet:2014jja,Duguet:2015yle} is nothing but $W(\varphi)-W_0(\varphi)$~\cite{qiu17a}. Eventually, both formalisms are thus structurally identical. However, they effectively differ in the way the disentangled clusters are actually determined as is illustrated in details in the next section. The latter difference is of importance in a practical, i.e. truncated, implementation of the formalism as is illustrated later via the numerical application.

\subsection{Amplitude equations}
\label{ampli}

Before coming to the disentangled clusters, let us first make a comment on the determination of the gauge-unrotated ones. In the PBCC context, the $U_n$ operators may be determined in a projection-after-optimization (PAO) scheme, i.e. they can be obtained in a first step by solving standard BCC equations (Eq.~\ref{angleindepBCCequat}), or via a more involved optimization-after-projection (OAP) scheme by solving BCC equations in  presence of the projection operator~\cite{Qiu2018}, e.g.
\begin{subequations}
\label{angleindepPBCCequat}
\begin{align}
0&= \langle \Phi^{k_{1}k_{2}} | P^\textrm{A} \, (H - E^\mathrm{A}) \,  e^{U} | \Phi \rangle \, ,
\\
0&= \langle \Phi^{k_{1}k_{2}k_{3}k_{4}} | P^\textrm{A} \, (H - E^\mathrm{A}) \,  e^{U} | \Phi \rangle \, .
\end{align}
\end{subequations}
The distinction between PAO and OAP is similar in spirit to obtaining the Bogoliubov transformation from a variational problem in presence or in absence of $P^\textrm{A}$ as discussed earlier.

Let us now come to the main point concerning the disentangled clusters. As for the gauge-dependent constant $W_0(\varphi)$, the same first-order ordinary differential equation (ODE) was proposed in Refs.~\cite{Duguet:2014jja,Duguet:2015yle} and~\cite{qiu17a}  (see Sec.~\ref{ampli2} below). Contrarily, the equations used to determine the gauge-rotated cluster operator ${\cal T}(\varphi)=W(\varphi)-W_0(\varphi)$ are different in both variants of the formalism. While amplitude equations generalizing those at play in BCC, i.e. at $\varphi=0$, were put forward in Ref.~\cite{Duguet:2014jja,Duguet:2015yle}, a set of first-order ODEs were introduced in Ref.~\cite{qiu17a}.  In the limit where PBCC becomes exact, the two versions of the formalism are identical. When PBCC is not exact, however, solving angle-dependent BCC amplitude equations for ${\cal T}(\varphi)$ produces a constant connected hamiltonian kernel $h(\varphi)$, as a consequence of which PBCC implemented in this manner is in fact energetically equivalent to BCC and must be discarded. See App.~\ref{AppEthan} for more details on this important point. 

Let us now detail the two options to determine the disentangled cluster amplitudes.

\subsubsection{Angle-dependent BCC equations}
\label{ampli1}

Left-multiplying the Schr\"odinger equation by the $n$-tuply (i.e. $2n$ quasiparticle) excited rotated Bogoliubov state
\begin{align}
\langle \tilde{\Phi}^{\mu}(\varphi) | \equiv \langle \Phi(\varphi) |  {\cal B}_{\mu} = \langle \Phi | R(\varphi) {\cal B}_{\mu} \, , \label{ntuplyexcitedrotated}
\end{align}
one obtains the equations of motion 
\begin{align}
\langle \tilde{\Phi}^{\mu}(\varphi) | H e^{U} | \Phi \rangle &=  E^{\text{A}} \, \langle \tilde{\Phi}^{\mu}(\varphi) | \, e^{U} | \Phi \rangle \, , \label{equatmotionenergy1}
\end{align}
which can be written in a compact form by invoking n-tuply excited norm and Hamiltonian kernels\footnote{The off-diagonal kernels introduced in Eq.~\ref{excitkernelsB} are those in use in Refs.~\cite{Duguet:2014jja,Duguet:2015yle}. As for Ref.~\cite{qiu17a}, they correspond to the "auxiliary" kernels introduced in Eq.~43. They constitute the key kernels from which both types of equations for the amplitudes discuss below derive naturally.} defined through
\begin{subequations}
\label{excitkernelsB}
\begin{align}
\tilde{{\cal N}}_{\mu}(\varphi)  &\equiv \langle \tilde{\Phi}^{\mu}(\varphi) | e^{U} | \Phi \rangle \nonumber \\
&=  \langle \Phi(\varphi) | {\cal B}_{\mu} \, e^{U} | \Phi \rangle \nonumber \\
&= {\cal N}(\varphi) \langle \Phi^{\mu} |  \, e^{{\cal T}(\varphi)} | \Phi \rangle  \, , \label{excitkernelsB1} \\
\tilde{{\cal H}}_{\mu}(\varphi) &\equiv \langle \tilde{\Phi}^{\mu}(\varphi) |  H e^{U} | \Phi \rangle \nonumber\\
&= \langle \Phi(\varphi) | {\cal B}_{\mu} \, H e^{U} | \Phi \rangle \nonumber \\
&= {\cal N}(\varphi) \langle \Phi^{\mu} |  H_{Z}(\varphi) \, e^{{\cal T}(\varphi)} | \Phi \rangle  \, . \label{excitkernelsB2} 
\end{align}
\end{subequations}
which reduce to the unexcited kernels introduced in Eq.~\ref{kernelsB} for ${\cal B}_{\mu}=1$.

After several steps of algebraic manipulations detailed in App.~\ref{angledependentequations}, Eq.~\ref{equatmotionenergy1} can be rewritten as a workable set of equations for the n-tuple gauge-dependent cluster amplitudes. Illustrating this for single and double excitations, Eq.~\ref{equatmotionenergy1} reduces to
\begin{subequations}
\label{angledepBCCequat}
\begin{align}
0&= \langle \Phi^{k_{1}k_{2}} | H_{Z}(\varphi)  e^{{\cal T}(\varphi)} | \Phi \rangle_{C}  \, , \label{angledepBCCequat1} \\
0&= \langle \Phi^{k_{1}k_{2}k_{3}k_{4}} | H_{Z}(\varphi)  e^{{\cal T}(\varphi)} | \Phi \rangle_{C}  \, , \label{angledepBCCequat2} 
\end{align}
\end{subequations}
in which the exponential naturally terminates by virtue of the connected character of the kernel. Appendix \ref{angledependentequations} and Ref. \onlinecite{Duguet:2015yle} contain a fuller derivation of these equations, but the basic idea is that the excited Hamiltonian kernels factor into a sum of products of \textit{connected} Hamiltonian kernels multiplied by excited norm kernels, and satisfaction of the Schr\"odinger equation implies that the excited connected Hamiltonian kernels vanish, which is precisely the contents of Eqs.~\ref{angledepBCCequat}.

Equations~\ref{angledepBCCequat} are formally identical to standard BCC equations except that the Hamiltonian $H$ is replaced by its similarity-transformed partner $H_{Z}(\varphi)$ at each gauge angle $\varphi$. As a matter of fact,  Eq.~\ref{angledepBCCequat} reduces at $\varphi=0$ to plain, i.e. unrotated, BCC equations (Eq.~\ref{Eqn:BCCAmp}) that are to be used to determine the plain cluster amplitudes $U_n$, e.g.
\begin{subequations}
\label{angleindepBCCequat}
\begin{align}
0&= \langle \Phi^{k_{1}k_{2}} | H  e^{U} | \Phi \rangle_{C}  \, , \label{angleindepBCCequat1} \\
0&= \langle \Phi^{k_{1}k_{2}k_{3}k_{4}} | H  e^{U} | \Phi \rangle_{C}  \, . \label{angleindepBCCequat2} 
\end{align}
\end{subequations}
Determining the angle-dependent cluster amplitudes ${\cal T}_1(\varphi)$, ${\cal T}_2(\varphi)$, \ldots by solving the angle-dependent BCC equations (Eq.~\ref{angledepBCCequat}) was put forward in Refs.~\cite{Duguet:2014jja,Duguet:2015yle}. Gauge-rotated BCC amplitude equations do not provide access to the norm kernel and must thus be complemented by another equation delivering $W_0(\varphi)$.  To do so, it was proposed in Refs.~\cite{Duguet:2014jja,Duguet:2015yle} to use a first-order ordinary differential equation (ODE) that belongs in fact to a systematic set of first-order ODEs that can be used to determine all $W_n(\varphi)$.

\subsubsection{Ordinary differential equations}
\label{ampli2}

At gauge angle $\varphi = 0$, the rotation operator $R(\varphi)$ is the identity.  Accordingly $U_Z(\varphi=0) = U$, and therefore
\begin{subequations}
\label{initialconditions} 
\begin{align}
W_0(0) &= 0,
 \label{initialconditions1}
\\
W_{k}(0) &= U_{k}.
\label{initialconditions2}
\end{align}
\end{subequations}
That we know a set of initial conditions suggests that one might obtain $W(\varphi)$ by some form of evolution.  The differential equations governing $W(\varphi)$ can be obtained\footnote{See App.~\ref{diffequations} for an alternative derivation.} by differentiating both sides of Eq.~\ref{WDef} defining $W(\varphi)$.  Writing Eq.~\ref{WDef} as
\begin{equation}
e^{Z(\varphi)} \, e^U |\Phi\rangle = e^{W(\varphi)} |\Phi\rangle
\end{equation}
and differentiating it, one finds 
\begin{equation}
\frac{\mathrm{d} Z(\varphi)}{\mathrm{d}\varphi} \, e^{Z(\varphi)} \, e^U | \Phi \rangle = \frac{\mathrm{d}W(\varphi)}{\mathrm{d}\varphi} \, e^{W(\varphi)} | \Phi \rangle
\end{equation}
or, since $W(\varphi)$ is in normal order and commutes with its derivative,
\begin{equation}
\frac{\mathrm{d}W(\varphi)}{\mathrm{d}\varphi} | \Phi \rangle = e^{-W(\varphi)} \, \frac{\mathrm{d} Z(\varphi)}{\mathrm{d}\varphi} \, e^{W(\varphi)} | \Phi \rangle.
\label{Eqn:WDeriv}
\end{equation}
It can be shown that\footnote{The operator $iA^{02}_{Z}(\varphi)$ is nothing but the operator $X$ of Ref.~\cite{qiu17a}.}
\begin{equation}
\frac{\mathrm{d} Z(\varphi)}{\mathrm{d}\varphi} = i \, A^{02}_Z(\varphi),
\end{equation}
where $A^{02}_Z(\varphi)$ is the de-excitation part of the similarity-transformed number operator
\begin{equation}
A_Z(\varphi) \equiv e^{Z(\varphi)} \, A \, e^{-Z(\varphi)}.
\end{equation}
Left-multiplying Eq.~\ref{Eqn:WDeriv} by all possible $\langle \Phi| \mathcal{B}^\mu$ leaves us with a set of coupled ODEs for the amplitudes defining $W_n(\varphi)$ for $n\geq 0$.  For example, this gives up to $W_{2}(\varphi)$
\begin{align}
\frac{d }{d\varphi} W_{0}(\varphi) &= \frac{i}{2} \sum_{k_1 k_2}  A^{02}_{k_1 k_2}(\varphi)W_{k_1 k_2}(\varphi)   \, , \label{ntuply1storderdiffeq0} \\
\frac{d }{d\varphi} W_{k_1k_2}(\varphi) &=  i \sum_{k_3 k_4} A^{02}_{k_3 k_4}(\varphi) \Big[ \frac{1}{2}W_{k_3 k_4k_1 k_2 }(\varphi) \nonumber \\
&\hspace{2.5cm} - W_{k_1 k_3}(\varphi)W_{k_2 k_4}(\varphi) \Big]   \, , \nonumber \\
\frac{d }{d\varphi} W_{k_1k_2k_3k_4}(\varphi) &=  i \sum_{k_5 k_6} A^{02}_{k_5 k_6}(\varphi) \big[ \frac{1}{2}W_{k_5k_6k_1  k_2 k_3 k_4}(\varphi) \nonumber \\
&\hspace{2.5cm} +W_{k_1 k_5}(\varphi) W_{k_6 k_2 k_3 k_4}(\varphi) \nonumber \\
&\hspace{2.5cm} + W_{k_2 k_5}(\varphi) W_{k_1 k_6 k_3 k_4}(\varphi) \nonumber \\
&\hspace{2.5cm}  + W_{k_3 k_5}(\varphi) W_{k_1 k_2 k_6 k_4}(\varphi) \nonumber \\
&\hspace{2.5cm}  + W_{k_4 k_5}(\varphi) W_{k_1 k_2 k_3 k_6}(\varphi) \big] \nonumber \, .
\end{align}
The above  ODEs stipulate that the dependence of the cluster amplitudes on the gauge angle $\varphi$ is driven by kernels of the particle number operator (i.e. of its de-excitation part) that is nothing but the infinitesimal generator of the $U(1)$ group. From the above examples, it is clear that the derivative of $W_k(\varphi)$ involves contributions from $W_{k+1}(\varphi)$.  By truncating at a certain excitation level, these equations can be decoupled, and the low order amplitudes can be obtained from an ODE solver, supplemented by the initial conditions given in Eq.~\ref{initialconditions}. Note however that truncating the set of ODEs (for any given truncation of the cluster operator $U$) implies an approximation on the action of the projection operator. The effect of this truncation will be gauged in the numerical application below.

In Refs.~\cite{Duguet:2014jja,Duguet:2015yle}, it was advocated to complement Eq.~\ref{angledepBCCequat} with the first differential equation above to access $W_{0}(\varphi)$. In Ref.~\cite{qiu17a}, it was advocated to determine angle-dependent cluster amplitudes $W_n(\varphi)$ by solving the set of differential equations.

While the derivation of the ODEs given above is straightforward, it may be useful to provide an alternative perspective providing a deeper insight. Because projected states are eigenstates of the particle number operator $A$, one may write
\begin{align}
\langle \Phi | A P^{\text{A}} {\cal B}_{\mu} e^{U}|\Phi\rangle = \text{A} \langle \Phi | P^{A} {\cal B}_{\mu} e^{U}|\Phi\rangle,
\label{physics1}
\end{align}
which can be rearranged to yield\footnote{When the de-excitation operator is trivial, i.e. ${\cal B}_\mu=1$, Eq.~\ref{physics1} is nothing but the left projection of the eigenequation for the particle number
\begin{align}
A | \Psi^{\text{A}} \rangle &=  \text{A} \, | \Psi^{\text{A}} \rangle \, , \label{schroedA} 
\end{align}
by the reference bra $\langle \Phi |$.}
\begin{align}
\text{A} &= \frac{\int_{0}^{2\pi} \!d\varphi \, e^{-i\text{A}\varphi}  \, \breve{{\cal A}}_{\mu}(\varphi)}{\int_{0}^{2\pi} \!d\varphi \, e^{-i\text{A}\varphi}  \, \tilde{{\cal N}}_{\mu}(\varphi)}   \, , \label{physics2}
\end{align}
where the n-tuply excited particle-number kernel\footnote{The nature of the operator kernel $\breve{{\cal A}}_{\mu}(\varphi)$ differs from the one introduced in Eq.~\ref{excitkernelsB2} for the Hamiltonian by the relative position of the de-excitation operator ${\cal B}_{\mu}$ and of the operator of interest, i.e. $H$ or $A$. Most interestingly, while an equation similar to Eq.~\ref{equatmotionenergy1} could be written for $A$, the same is not true regarding Eq.~\ref{physics1} for $H$.} involved is defined as
\begin{align}
\breve{{\cal A}}_{\mu}(\varphi) &\equiv \langle \Phi(\varphi) |  A \, {\cal B}_{\mu} \, e^{U} | \Phi \rangle\, . \label{newkernelsA}
\end{align}

Equation~\ref{physics2} provides a set of exact identities highlighting the effect of the symmetry projection when focusing on the particle number operator {\it for which we know the exact answer}, i.e. the number $\text{A}$. Contrarily to the gauge-rotated BCC equations (Eq.~\ref{angledepBCCequat}) deriving from the Schr\"odinger equation, these identities reflect the structure of the $U(1)$ symmetry group and are independent of the actual dynamics generated by the Hamiltonian. While Eq.~\ref{angledepBCCequat} equates the BCC residuals to zero at each gauge angle $\varphi$, property~\ref{physics2} spans the entire volume of the symmetry group at once.  As a matter of fact, fulfilling Eq.~\ref{physics2} for each $\mu$ can be rephrased into satisfying a set of differential equations stipulating how the evolution of the n-tuply excited norm kernel with respect to the gauge angle is driven by the n-tuply excited particle-number kernel 
\begin{align}
\frac{d }{d\varphi} \tilde{{\cal N}}_{\mu}(\varphi) &= \frac{d }{d\varphi} \langle \Phi |  R(\varphi) {\cal B}_{\mu} \, e^{U} | \Phi \rangle  = i \breve{{\cal A}}_{\mu}(\varphi) \label{physics3}  \, .
\end{align}
Indeed, satisfying Eq.~\ref{physics3} implies that
\begin{align}
\frac{\int_{0}^{2\pi} \!d\varphi \, e^{-i\text{A}\varphi}  \, \breve{{\cal A}}_{\mu}(\varphi)}{\int_{0}^{2\pi} \!d\varphi \, e^{-i\text{A}\varphi}  \, \tilde{{\cal N}}_{\mu}(\varphi)} &= -i \frac{\int_{0}^{2\pi} \!d\varphi \, e^{-i\text{A}\varphi}  \, \frac{d }{d\varphi} \tilde{{\cal N}}_{\mu}(\varphi)}{\int_{0}^{2\pi} \!d\varphi \, e^{-i\text{A}\varphi}  \, \tilde{{\cal N}}_{\mu}(\varphi)} \nonumber \\
&= +i \frac{\int_{0}^{2\pi} \!d\varphi \, [\frac{d }{d\varphi} e^{-i\text{A}\varphi}]  \,  \tilde{{\cal N}}_{\mu}(\varphi)}{\int_{0}^{2\pi} \!d\varphi \, e^{-i\text{A}\varphi}  \, \tilde{{\cal N}}_{\mu}(\varphi)}  \nonumber \\
&= \text{A} \, , \nonumber
\end{align}
such that Eq.~\ref{physics2} is fulfilled. It happens that, after several steps of algebraic manipulations detailed in App.~\ref{diffequations}, Eq.~\ref{physics3} actually delivers the set of ODEs examplified in Eq.~\ref{ntuply1storderdiffeq0} from which the disentangled cluster amplitudes $W_n(\varphi)$, $n \geq 0$, are determined. This ends demonstrating that the set of ODEs is nothing but a rephrasing of Eq.~\ref{physics1}, which itself translates the restoration of good particle number symmetry at each n-tuply excitation level.

\section{Application}
\label{Sec:application}

We apply the PBCC formalism laid out in Sec.~\ref{Sec:many-body} to the pairing Hamiltonian problem at the single and double excitation levels (i.e. $U = U_1 + U_2$). In this context, the general Bogoliubov reference state simplifies into a simpler BCS vacuum. Consequently, the presently applied method is termed PBCS-CCSD, or PBCS-CCD in case one only includes doubles. Our main objective is to gauge the merits of the symmetry projection, which will be done by comparing PBCS-CCSD results to exact ones~\cite{Richardson1,Richardson4} and to those obtained in Ref.~\cite{Henderson:2014vka} via the BCS-CC(S)D method from which the particle-number projection is absent. The reader interested in comparing the results displayed below with those obtained with recently proposed alternatives of similar quality is referred to Refs.~\cite{Deg16,ripoche17a}. 

\begin{figure*}[t]
\includegraphics[width=0.96\columnwidth]{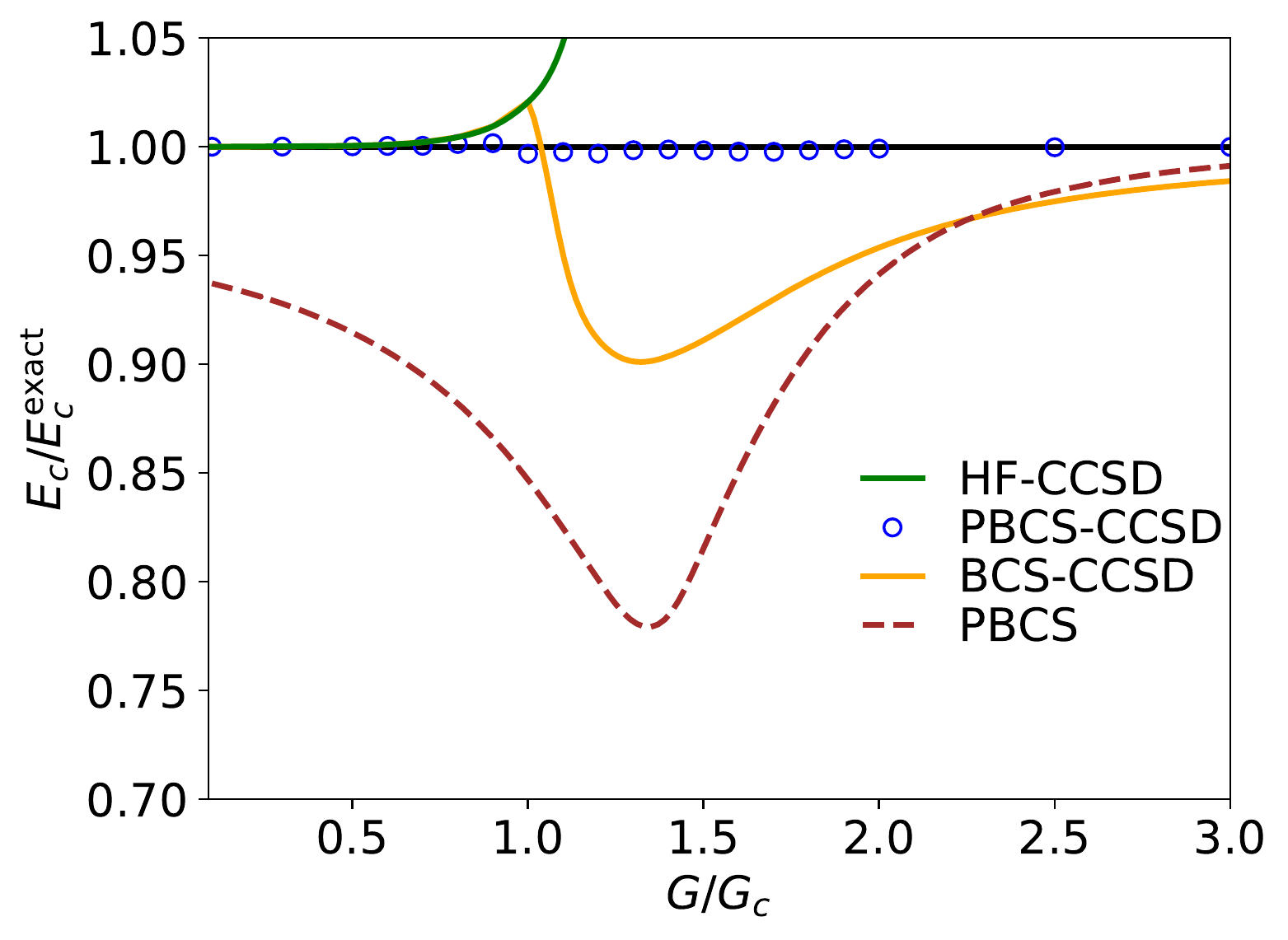}
\hfill
\includegraphics[width=0.96\columnwidth]{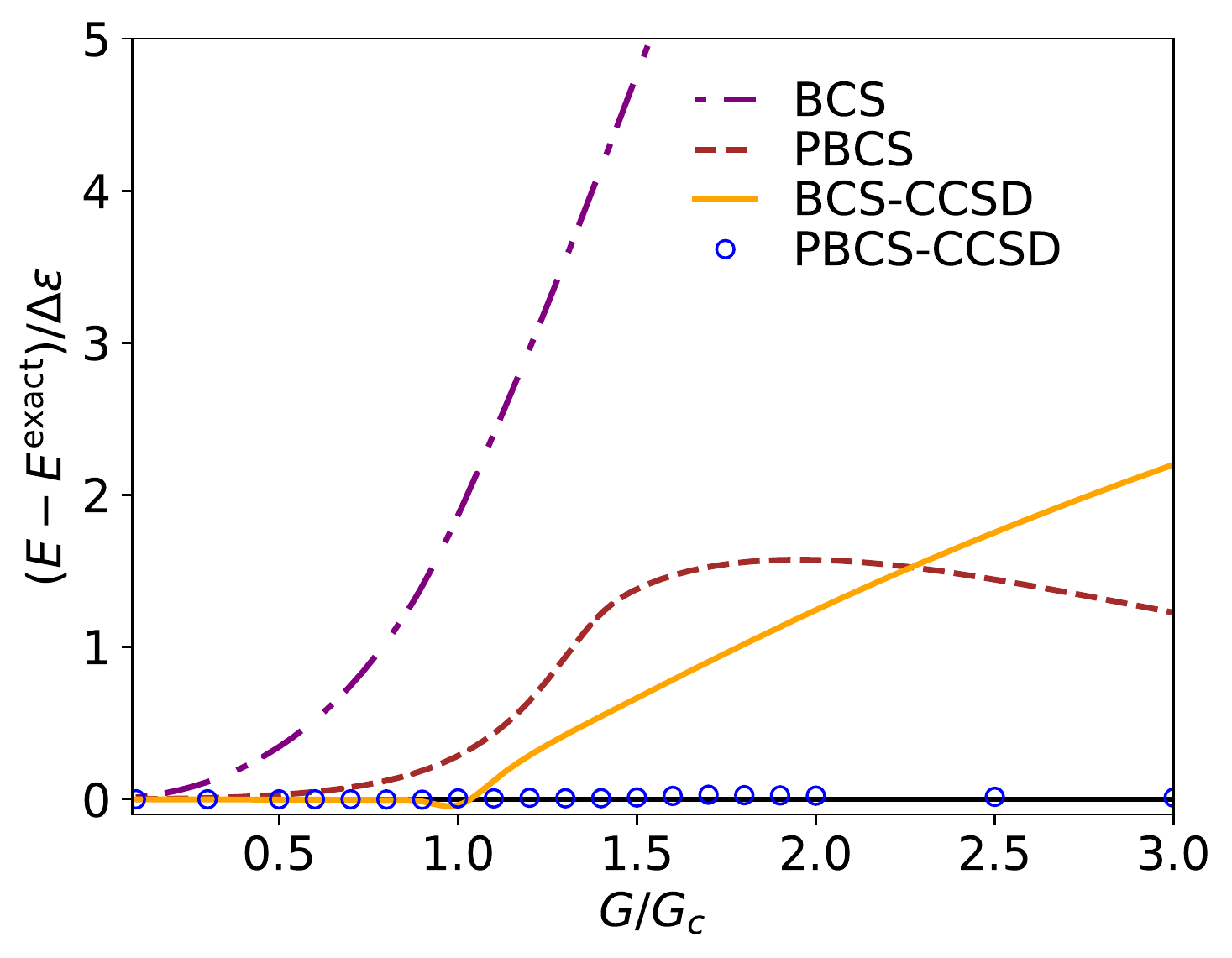}
\caption{(Color online) Energies for 100 levels at half filling. Left panel: Fraction of the correlation energy recovered. Right panel: Energy errors in units of $\Delta \epsilon$. 
\label{Fig:Ene}}
\end{figure*}

Results are only provided for the ODE-based approach since, as mentioned earlier and as discussed\footnote{Note that App.~\ref{AppEthan} also provides illustrative results regarding the behavior of the connected part of the hamiltonian kernel $h(\varphi)$ and of the norm kernel ${\cal N}(\varphi)$ that constitute the building blocks to compute the PBCC energy through Eq.~\ref{projeigenequatkernelsB1}.} in details in App.~\ref{AppEthan}, solving angle-dependent BCC equations for ${\cal T}(\varphi)$ makes PBCC energetically equivalent to BCC and must thus be discarded. In addition to the truncation of the cluster operator $U$ to singles and doubles, the set of ODEs is truncated at the doubles level, i.e. they are solved for $W_0(\varphi)$, $W_1(\varphi)$ and $W_2(\varphi)$ while assuming that $W_n(\varphi)=0$ for $n \geq 3$.

\begin{figure}[t]
\includegraphics[width=0.96\columnwidth]{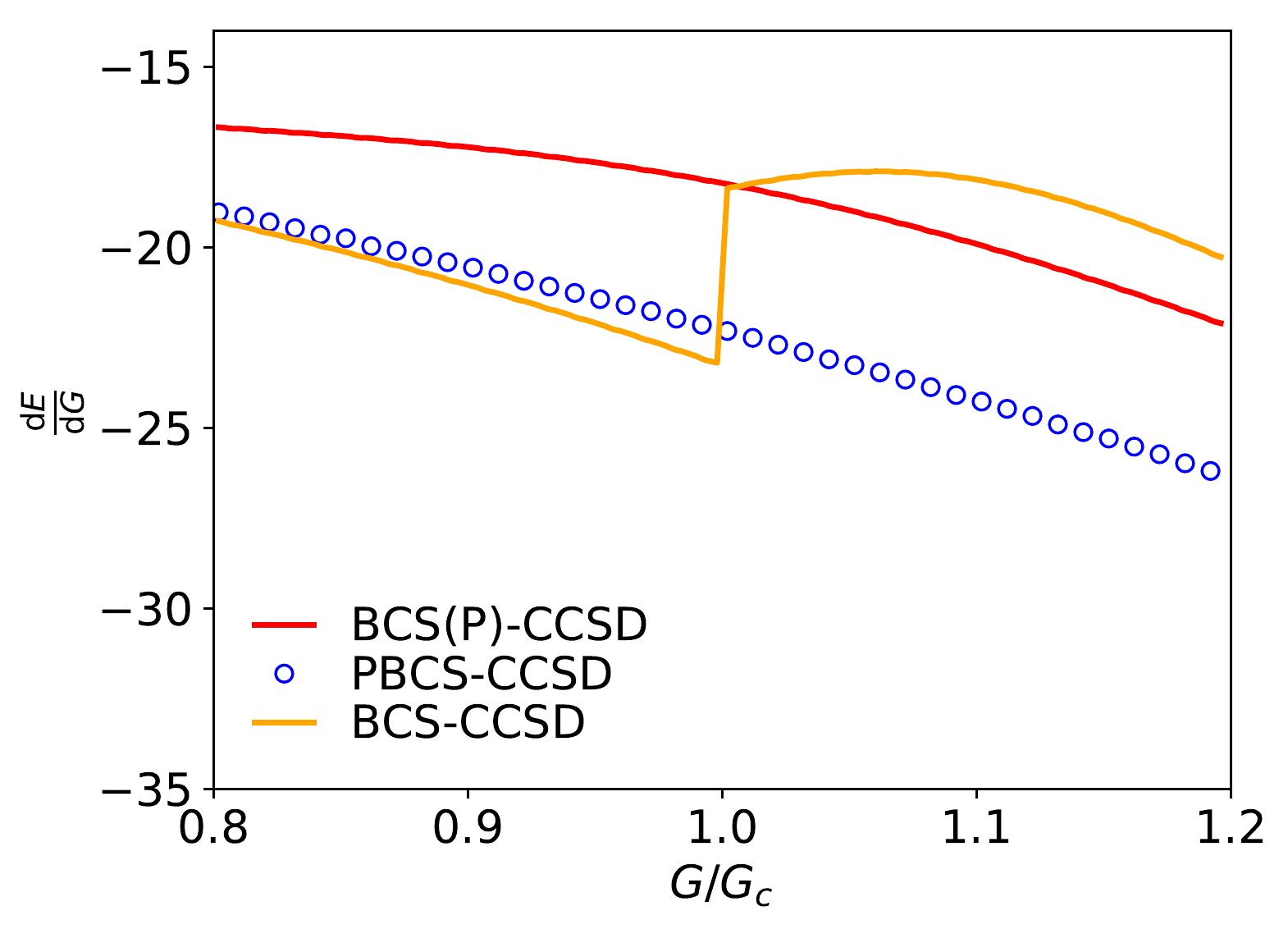}
\caption{Numerical derivative of the energy with respect to $G$ for 12 levels at half filling. Contrarily to BCS-CCSD, the calculation coined as BCS(P)-CCSD relies on the BCS reference state extracted from the VAP PBCS calculation and not on the plain variational BCS state.
\label{fig:derivative}}
\end{figure}

\subsection{Pairing Hamiltonian}
\label{Sec:Hamiltonian}

The pairing or reduced BCS Hamiltonian is defined as
\begin{equation}
H = \sum_p \left(\epsilon_p - \lambda\right) N_p - G \sum_{pq} P_p^\dagger \, P_q \, .
\label{Eqn:PairHam}
\end{equation}
Here, $\lambda$ is the chemical potential, $\epsilon_p$ denotes single-particle energy levels, and $G$ is the interaction strength.  The pair operators can be mapped to fermions using
\begin{subequations}
\begin{align}
N_p &= c_{p}^\dagger \, c_{p}^{} + c_{\bar{p}}^\dagger \, c_{\bar{p}}^{} \, ,
\\
P_p^\dagger &= c_{p}^\dagger \, c_{\bar{p}}^\dagger \, ,
\end{align}
\label{Eqn:DefNPhys}
\end{subequations}
and satisfy an $SU(2)$ algebra
\begin{subequations}
\begin{align}
[P_p,P_q^\dagger] &= \delta_{pq} \, \left(1 - N_p\right) \, ,
\\
[N_p,P_q^\dagger] &=  2\, \delta_{pq} \, P_q^\dagger \, .
\end{align}
\label{SU2}
\end{subequations}
Single-particle states $p$ and $\bar{p}$ are said to be pair conjugated and are degenerate such that $\epsilon_p =\epsilon_{\bar{p}}$. For simplicity, the single-particle levels are taken to be equally spaced, so that $\epsilon_p  = p \, \Delta \epsilon$ where $\Delta \epsilon$ is the level spacing.  Because the interaction is attractive, the mean-field solution spontaneously breaks $U(1)$ global-gauge symmetry for all $G$ greater than some critical value denoted as $G_c$ and takes the form of a BCS reference state. For $G < G_c$, the mean-field solution reduces to the symmetry conserving HF Slater determinant.

Having introduced the pairing Hamiltonian, the explicit expression of all the quantities necessary to implement PBCS-CCSD theory are provided in App.~\ref{PH}.

\subsection{Results}
\label{Sec:Results}

\subsubsection{Energetics}
\label{Sec:Energetics}

Figure \ref{Fig:Ene} displays (i) the fraction of correlation energy recovered with respect to the Hartree-Fock ground state in the left panel and (ii) the error in the total energy in the right panel. Results are shown  for 100 levels at half filling as a function of the interaction strength.

In the weakly correlated regime ($G \lesssim G_c$) the symmetry adapted coupled cluster (i.e. coupled cluster based upon a Hartree-Fock reference, here called HF-CCSD though the effects of singles are zero by symmetry) is fairly accurate. It is entirely equivalent to BCS-CCSD in this regime given that the BCS reference state reduces to the HF Slater determinant. Increasing $G$, HF-CCSD fails spectacularly when approaching $G_c$. Contrarily, BCS-CCSD continues to deliver reasonably accurate energies, albeit with a discontinuous derivative at $G = G_c$~\cite{Henderson:2014vka} reflecting the change in character of the mean-field reference state.  Number projected BCS (PBCS), happens to be wonderfully accurate for larger $G$ but, while exact at $G=0$, does not perform as well for small $G$.  Its error is largest for $G$ slightly larger than $G_c$. It is just a remarkable feature that PBCS is better for large $G$ and coupled cluster for small $G$ as well as that PBCS is better for smaller numbers of levels and coupled cluster for larger as the energetic benefit of number projection dwindles as the number of correlated particles increases.

\begin{figure}[t]
\includegraphics[width=\columnwidth]{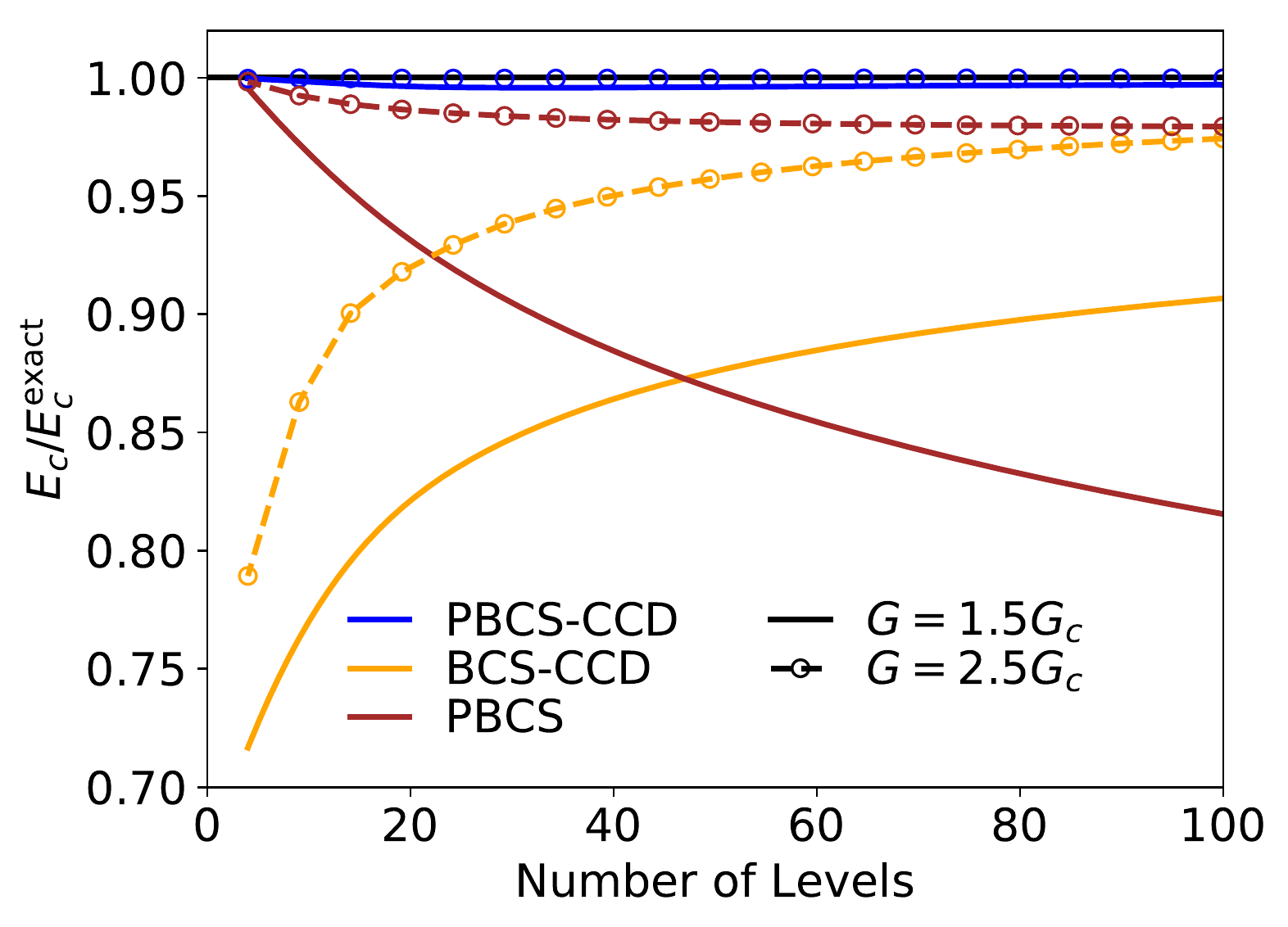}
\caption{(Color online) Fraction of the correlation energy recovered at half filling as a function of the number of levels. Two different values of $G/G_c$ are employed, knowing that the precise value of $G_c$ actually depends on the number of levels.
\label{Fig:EnergyNumberOfLevels}}
\end{figure}

While both PBCS and BCS-CCSD perform well for some values of $G$ and not as well for others, PBCS-CC gives almost exact energies for all $G$, even the region near $G_c$ where neither PBCS nor BCS-CCSD performs particularly well. As anticipated from the well-celebrated PBCS theory, the strongest impact of the projection is obtained in the regime where the symmetry is weakly broken in the first place, i.e. after the phase transition for intermediate coupling strengthes. Eventually, merging symmetry projection and coupled cluster theory leads thus to a wave function that is significantly better than either of its components.

Dealing with a finite quantum system, observables must behave smoothly through the normal-to-superfluid transition at $G_c$. This is indeed the case for exact solutions of the pairing Hamiltonian. As alluded to above, the BCS-CCSD energy does wrongly display a discontinuous derivative at $G_c$, which is in fact due to an abrupt change in character of the plain BCS reference state.  When the reference state is obtained in a VAP sense from PBCS, it evolves smoothly with $G$, and consequently so too do BCS-CC and PBCS-CC energies based upon it, as seen in Fig. \ref{fig:derivative}.  The first derivative discontinuity in the energy at $G = G_c$ is thus absent (present) when using the reference state from the PBCS (BCS) variational calculation, irrespective of whether the number projection operator acts on the BCS-CC wave function or not.

The fractional correlation picture can be misleading in certain cases.  The right panel shows that for small $G$, all methods are close to the right answer.  For larger $G$, the error in PBCS decreases (indeed, PBCS is an eigenfunction of the Hamiltonian in the $G \to \infty$ limit) while the total errors of BCS and BCS-CCD are still increasing at $G = 3 \, G_c$.  Adding number projection to BCS-CCD yields an absolute error in the total energy too small to be distinguished on the plot.

The benefits of symmetry projection are most pronounced for smaller systems, while as Ref. \cite{Henderson:2014vka} shows, BCS-CCD seems to be on average more accurate for larger system sizes.  It is thus worth checking how PBCS-CCD performs as a function of system size.  This is shown in Fig. \ref{Fig:EnergyNumberOfLevels} where the fraction of correlation energy recovered at half filling is displayed as a function of the number of levels for two different values of $G/G_c$.  In keeping with our expectations, PBCS tends to perform worse as system size increases, while BCS-CCD does better.  While the quality of PBCS-CCD does deteriorate ever so slightly with increasing system size, the method delivers nearly exact energies across the board.

\begin{figure}[t]
\includegraphics[width=\columnwidth]{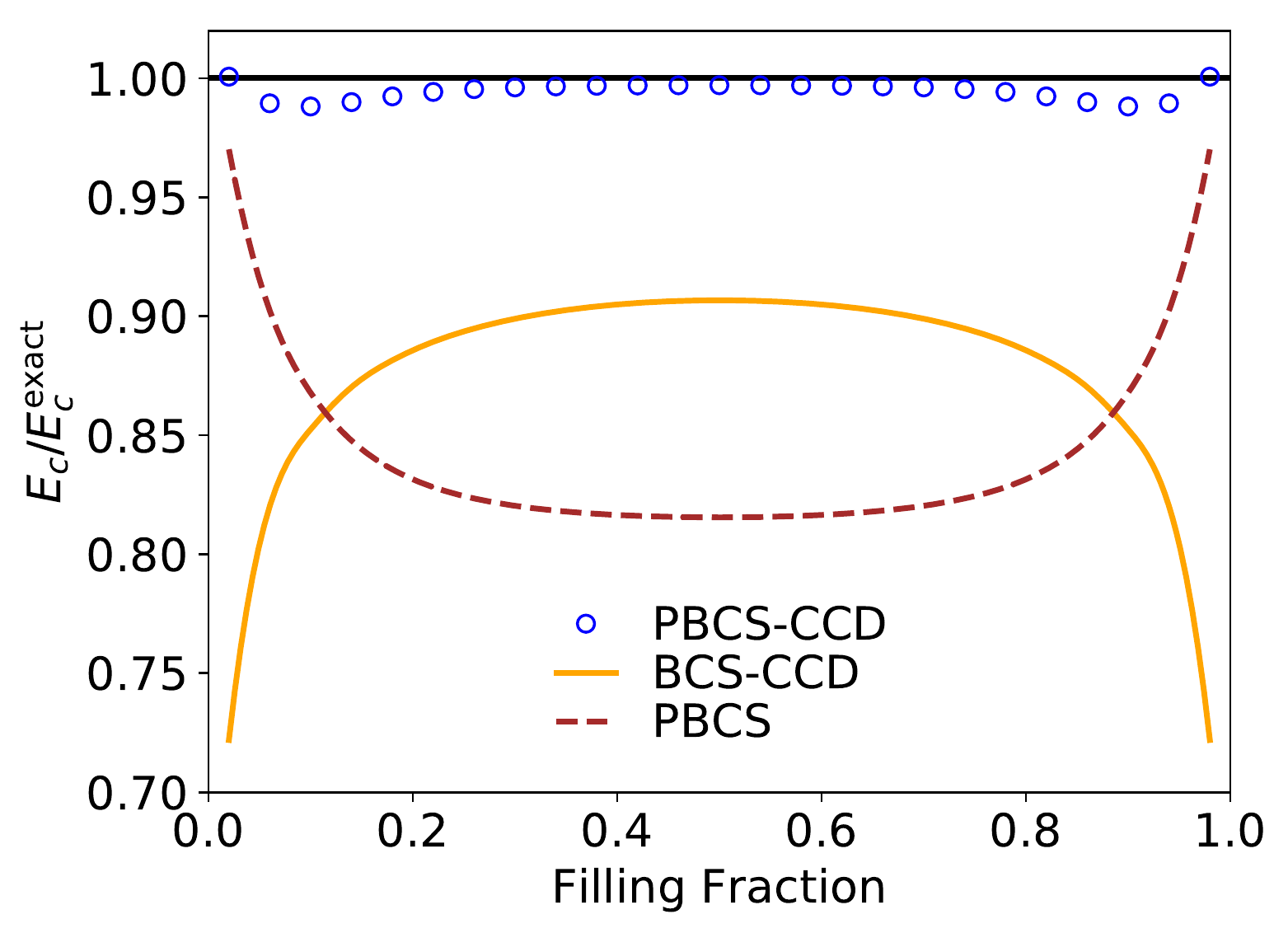}
\caption{(Color online) Fraction of correlation energy recovered for 100 levels as a function of the filling fraction. The interaction strength is $G = 1.5 \, G_c$, where here $G_c$ is the critical $G$ at half filling.
\label{Fig:FillingFraction}}
\end{figure}

The situation away from half filling is broadly similar, as can be seen in Fig. \ref{Fig:FillingFraction} where the fraction of correlation energy recovered for 100 levels as a function of the filling fraction is displayed. The interaction strength $G = 1.5 \, G_c$ is employed, where $G_c$ denotes here the critical $G$ at half filling.  This value of $G$ is sufficiently large that number symmetry is broken for all filling fractions under consideration.  BCS-CCD is most accurate near half filling, while PBCS is in contrast most accurate for small or large filling fractions.  PBCS-CCD seems to combine these two effects and is exceptionally accurate for all filling fractions. On the other hand, the strongest benefit from the projection is obtained for filling fractions close to zero or one, i.e. where the symmetry breaking is the weakiest\footnote{In the nuclear physics context, this typically corresponds to semi-magic nuclei near shell closures for which we expect PBCC to be the most superior to BCC. This however remains to be seen in realistic nuclear applications.}. Note that single excitations have been presently excluded from BCS-CC and PBCS-CC calculations simply because including them is more cumbersome computationally. Because single excitations have the effect of adjusting the reference determinant, BCS-CCSD or PBCS-CCSD results can only be expected to be better than those displayed in Fig~\ref{Fig:FillingFraction}.

\subsubsection{One-body observables}
\label{Sec:one-body}

Thus far, the focus has been on the energy. It is important to check the quality of the predictions for other properties as well. Generally, properties can be computed by differentiating the total energy with respect to some parameter in the Hamiltonian. For example, single-particle occupation probabilities $n_p$ can be expressed as $\mathrm{d}E/\mathrm{d}\epsilon_p$. From them, any one-body ground-state quantity can be obtained.

\begin{figure}[t]
\includegraphics[width=\columnwidth]{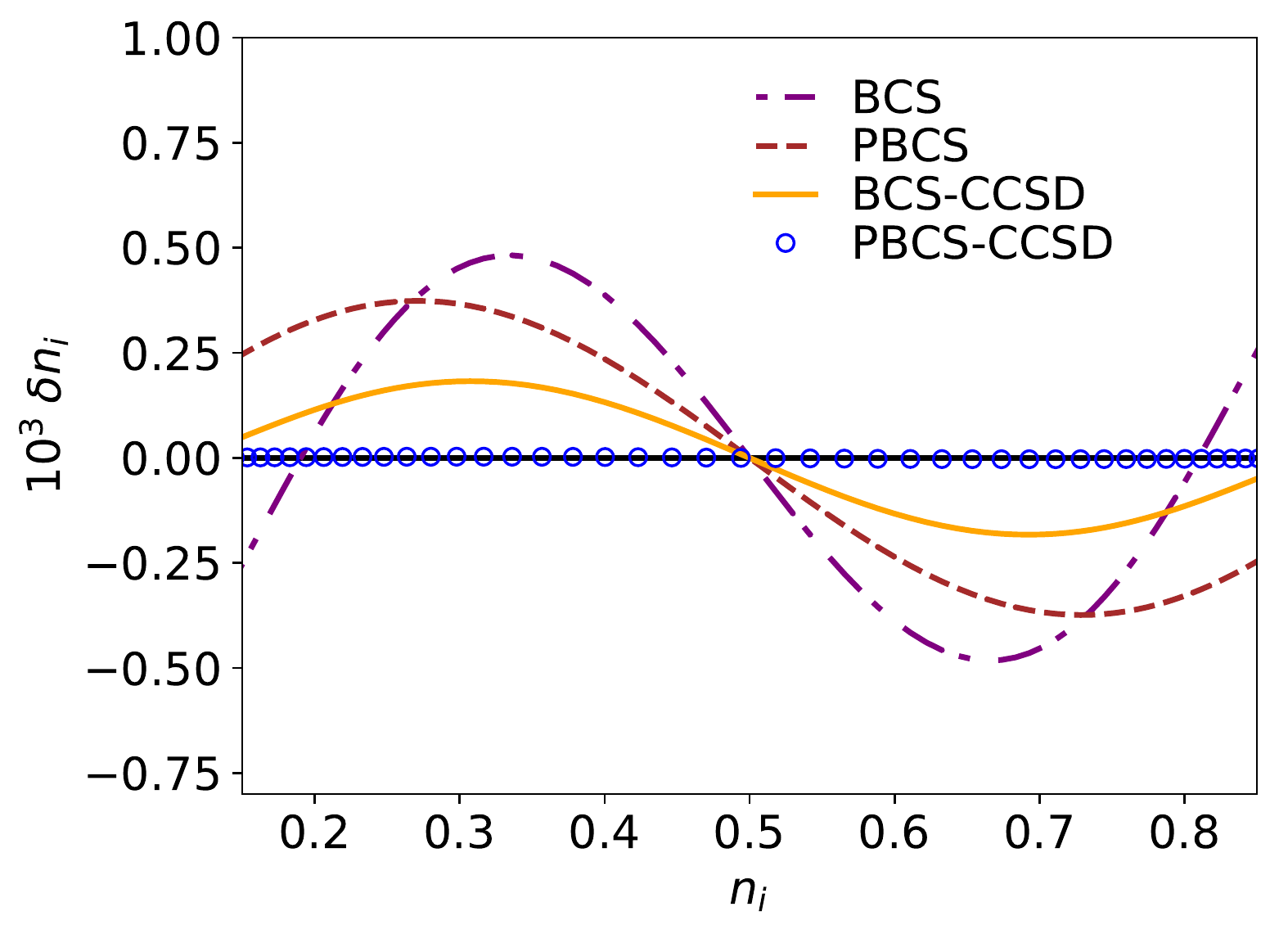}
\caption{(Color online) Deviations from exact occupation numbers for 100 levels at half filling. The interaction strength is $G/\Delta\epsilon = 1$ ($G/G_c = 5.5$).
\label{Fig:Occ}}
\end{figure}

The errors on single-particle occupation probabilities are shown in Fig. \ref{Fig:Occ}. As one might expect BCS displays noticeable errors that are to some extent mitigated by PBCS.  It was already noted in Ref.~\onlinecite{Henderson:2014vka} that BCS-CCD gives surprisingly accurate occupancies.  Adding number projection yields essentially no error such that PBCS-CCSD delivers basically exact results for any one-body property.

\subsubsection{Particle-number restoration}
\label{Sec:PNR}

\begin{figure}[t]
\includegraphics[width=\columnwidth]{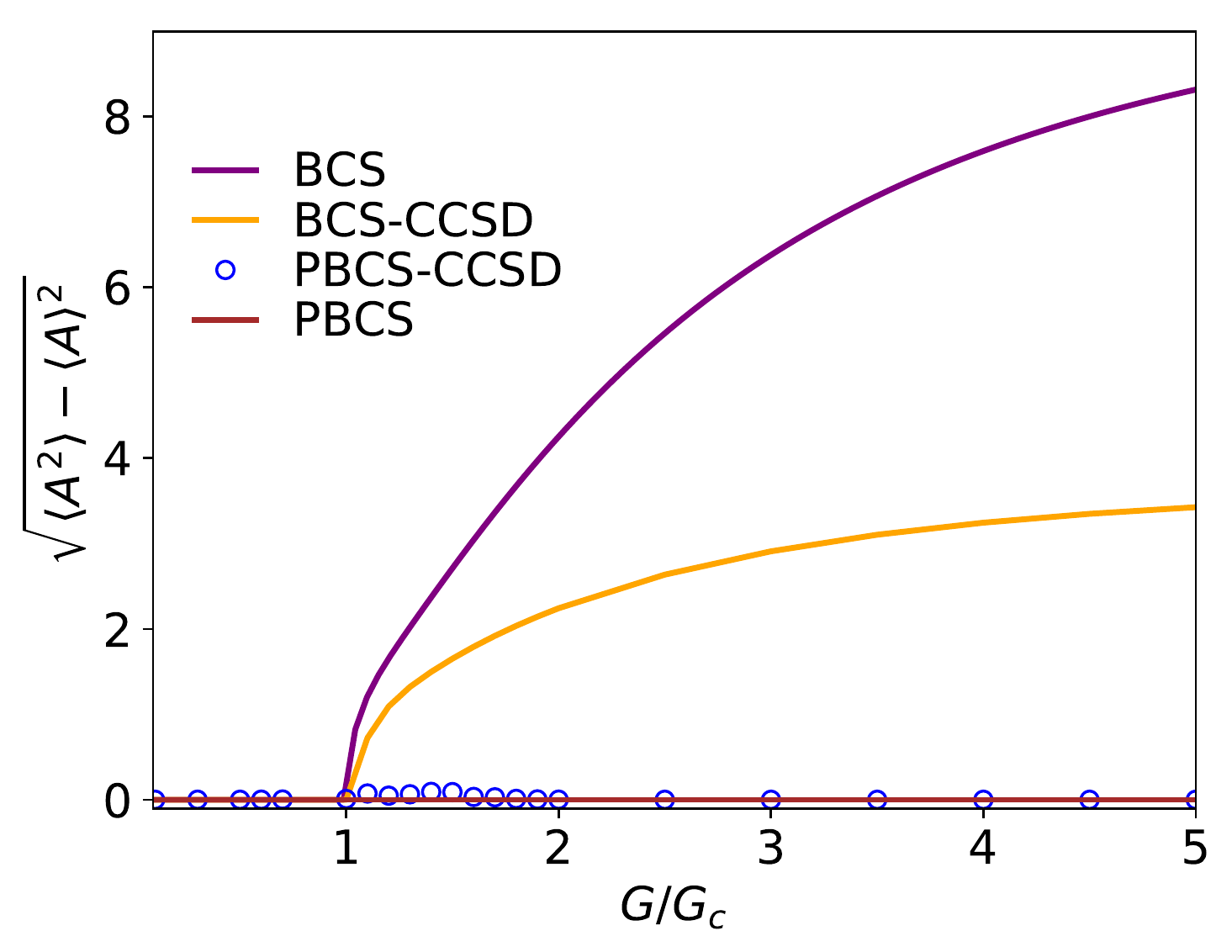}
\caption{(Color online) Dispersion of particle number for 100 levels at half filling as a function of $G/G_c$. Results are displayed for BCS, BCS-CCSD, PBCS and PBCS-CCSD.
\label{Fig:dispersion}}
\end{figure}

In addition to the great performance of PBCS-CC for ground-state energy and one-body observables, the actual restoration of good particle number at each BCC truncation order must be proven. To do so, Fig.~\ref{Fig:dispersion} displays the dispersion of particle number 
\begin{equation}
\sigma_{\text{A}}\equiv \sqrt{\langle A^2 \rangle - \langle A \rangle^2} \, , 
\end{equation}
with and without projection as well as with and without BCC corrections, i.e. at the (P)BCS and (P)BCS-CCSD levels.

As expected, one first observes that the particle-number dispersion in zero in all cases for $G<G_c$ given that the methods are symmetry-conserving to begin with over this interval. Beyond $G_c$, the dispersion grows monotonically in the BCS calculation, reaching large values in the strongly correlated regime. While including dynamical correlations at the BCS-CCSD level does effectively reduce the particle-number dispersion while improving the energetics, the dispersion remains significant\footnote{It remains of interest to test in which fashion the particle number is effectively restored as one includes higher cluster amplitudes toward the exact BCC wave function.}. On the other hand, including the projection on top of BCS does restore the symmetry exactly for all values of $G_c$ via the inclusion of non-dynamical correlations but misses dynamical correlations that are of importance for high-precision energetics as illustrated previously. Eventually, going to PBCS-CCSD brings the best of both worlds, i.e. both dynamical and non-dynamical correlations are captured in a consistent way such that the particle-number symmetry is indeed restored for all values of $G_c$ while obtaining nearly perfect energetics. 

While being convincingly close to zero for any practical purpose, one observes that the dispersion is not strictly null near $G_c$ in the PBCS-CCSD calculation. This relates to the fact that the set of ODEs is truncated (i.e.  $W_0(\varphi)$, $W_1(\varphi)$ and $W_2(\varphi)$ are presently retained), which implies an approximation on the action of the projection operator. As a matter of fact, the present result demonstrates that the truncation of the ODEs at the double level does provide a faithful account of the projection operator for all values of the coupling strength. If needed, one can anyway envision to include $W_3(\varphi)$ to reach very high precision.

\subsubsection{Dependence on the chemical potential}
\label{Sec:chem}

\begin{figure}[t]
\includegraphics[width=\columnwidth]{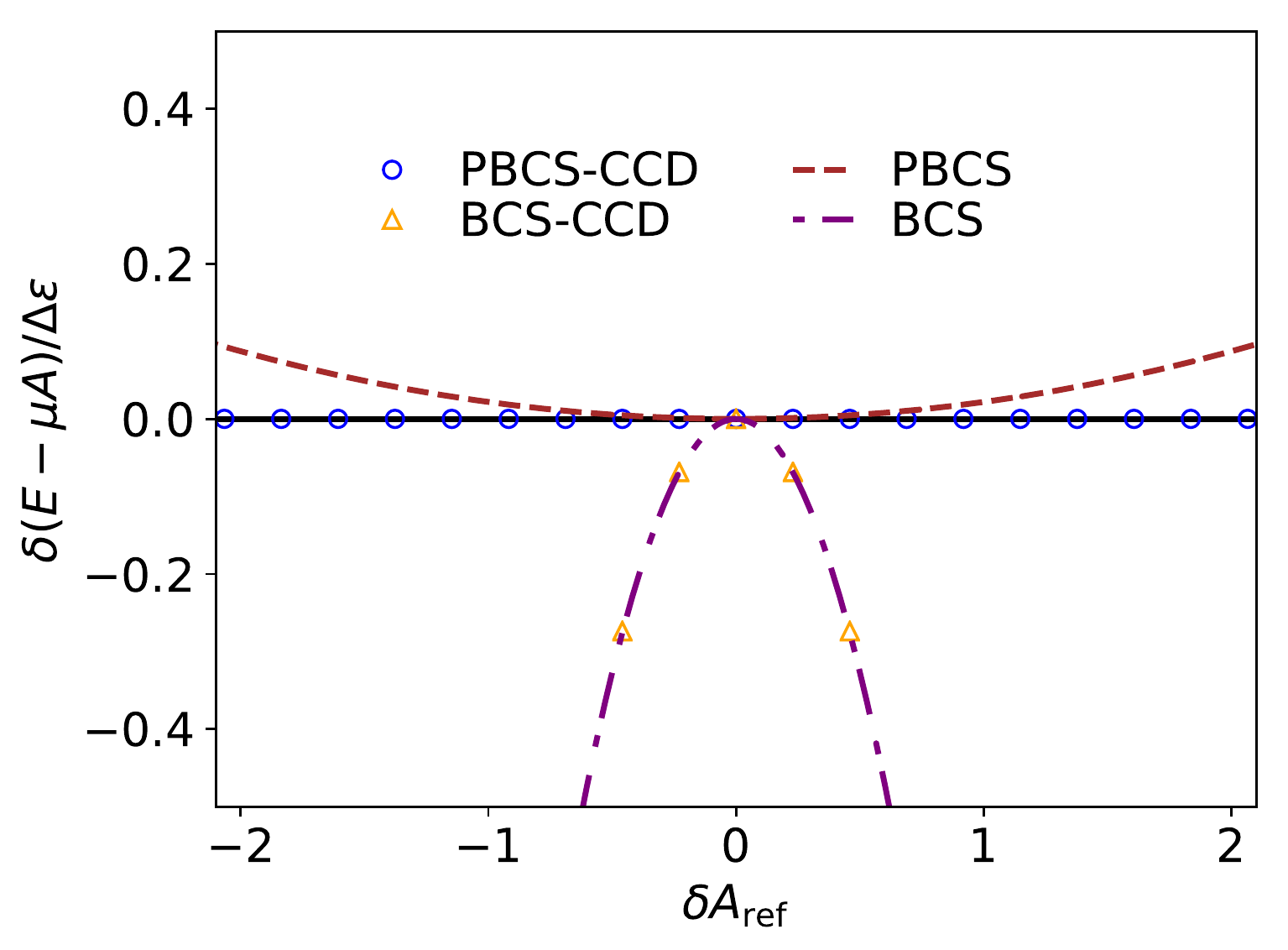}
\caption{(Color online) Energy variation with respect to the chemical potential of the underlying BCS reference state. Results are  in units of $\Delta \epsilon$ for 100 levels at half filling.  The interaction strength is $G/\Delta\epsilon = 1$ ($G/G_c = 5.5$).
\label{Fig:VaryChemPot}}
\end{figure}

At last, the dependence of the results on the chemical potential potential is investigated.  Both BCS and BCS-CC rely on a chemical potential to control the average number of particles. A question of interest is to what extent the PBCS-CC ground-state energy depends on the chemical potential $\lambda$ used to generate the underlying BCS reference state.  To this end, BCS, BCS-CCD, PBCS in a projection after variation (PAV) scheme, i.e. with $u$ and $v$ parameters fixed to their BCS values\footnote{In a VAP scheme, PBCS does not depend on the underlying chemical potential.}, and PBCS-CCD results are displayed in Fig. \ref{Fig:VaryChemPot} while varying the chemical potential around its BCS value. The quantity reported is the grand potential $\langle H - \lambda (A - \text{A}) \rangle$, where $\text{A}=100$ denotes the targeted number of particles. Clearly, BCS and BCS-CCD produce essentially identical deviations, which means that the BCS-CCD correlation energy depends only weakly on the chemical potential. Projecting the corresponding BCS states, the PBCS energy depends (mildly) on the chemical potential, i.e. on the average particle number defining the underlying BCS state. Doing the same on top of the BCS-CCD states, the PBCS-CCD energy is independent of the underlying chemical potential. This demonstrate that the consistent combination of dynamical correlations through the inclusion of singles and doubles and of non-dynamical correlations via the projection erase the memory of the underlying BCS state at rather low CC order.

\section{Conclusions}
\label{Sec:Conclusions}

Coupled cluster theory very efficiently captures weak correlations but, when based upon a symmetry-adapted reference determinant, tends to break down when the mean-field solution is strongly unstable toward symmetry breaking.  While this failure can be avoided by using a symmetry-breaking reference state, one pays the price in lost quantum numbers and potential inaccuracies in properties sensitive to the broken symmetry.  The choice between accurate energetics on the one-hand and good symmetries on the other is resolved at the mean-field level by symmmetry projection, and the same appears to be true at the correlated level.  While symmetry-projected mean-field methods do not typically describe weak correlations and symmetry-adapted coupled cluster methods do not generally describe strong correlations, the combination of symmetry projection and coupled cluster theory yields a wave function that can accurately describe both regimes.  Indeed, for the pairing Hamiltonian PBCS-CC is energetically exact for both small and large $G$ and even gives essentially exact results near $G_c$ where neither PBCS nor BCS-CC alone is adequate.  Moreover, the wave function of PBCS-CC appears to yield essentially exact one-body properties.  All of this can be obtained with a computational cost not too much larger than that of the underlying BCS-CC.  While we have not yet applied our techniques to more physical Hamiltonians, these early applications suggest that symmetry-projected coupled cluster theory has the potential to be an important part of the computational arsenal.

\section*{Acknowledgements}

The work at Rice University work was supported by the US National Science Foundation under Awards No. CHE-1462434 and CHE-1762320. G.E.S. is a Welch  Foundation Chair (No. C-0036). Implementation of PBCS-CC was facilitated by the drudge algebraic manipulator and gristmill code printer developed by Jinmo Zhao.  We thank Jorge Dukelsky for providing his PBCS code.

\appendix

\section{Similarity-transformed operators}
\label{SecSimilarity}

\subsection{Thouless transformation between vacua}
\label{SecThouless}

The gauge-rotated Bogoliubov state can be expressed via a non-unitary Thouless transformation of the unrotated one according to
\begin{equation}
\label{thoulessbetweenbothvacua}
\langle \Phi(\varphi) | \equiv \langle \Phi(\varphi) | \Phi \rangle \langle \Phi | e^{Z(\varphi)} \, ,
\end{equation}
where the one-body Thouless operator 
\begin{equation}
\label{thoulessoprot2}
Z(\varphi) \equiv \frac{1}{2} \sum_{k_1k_2} Z^{02}_{k_1k_2} (\varphi)  \beta_{k_2} \beta_{k_1} \, ,
\end{equation}
solely contains a pure de-excitation part.  The skew-symmetric Thouless matrix
\begin{equation}
\label{thoulessoprot}
Z^{20} (\varphi) \equiv  N(\varphi) M^{-1} (\varphi)  \, ,
\end{equation}
is expressed via the {\it transition} Bogoliubov transformation made out of the two matrices
\begin{subequations}
\label{transfobogotransition}
\begin{align}
M(\varphi) & \equiv e^{-i\varphi} U^{\dagger} U + e^{+i\varphi} V^{\dagger} V \, , \label{transfobogotransitionM}  \\
N(\varphi) & \equiv e^{+i\varphi} U^{T} V + e^{-i\varphi} V^{T} U \, , \label{transfobogotransitionN} 
\end{align}
\end{subequations}
where $(U,V)$ define the unrotated Bogoliubov vacuum $| \Phi \rangle$.

\subsection{Similarity transformation}
\label{Sectransforop}

Similarity-transformed quasiparticle operators are defined through
\begin{align}
\beta^{(\dagger)}_{k}(\varphi) &\equiv e^{Z(\varphi)}\beta^{(\dagger)}_{k}e^{-Z(\varphi)} \label{transforqptilde1} \\
&= \beta^{(\dagger)}_{k} + [Z(\varphi),\beta^{(\dagger)}_{k}] + \frac{1}{2!} [Z(\varphi),[Z(\varphi),\beta^{(\dagger)}_{k}]] + \ldots\,  \nonumber
\end{align}
Given the elementary commutators
\begin{subequations}
\begin{align}
 \Big[ \beta^+_{k} \beta^+_{k'} , \beta_{k_1} \Big]   &= \beta^+_{k} \delta_{k' k_1} - \beta^+_{k'} \delta_{k k_1} \, , \\
 \Big[ \beta^+_{k} \beta^+_{k'} , \beta^+_{k_1} \Big] &= 0 \, , \\
 \Big[ \beta_{k} \beta_{k'} , \beta^+_{k_1} \Big] &= \beta_{k} \delta_{k' k_1} - \beta_{k'} \delta_{k k_1} \, , \\
 \Big[ \beta_{k} \beta_{k'} , \beta_{k_1} \Big]       &= 0 \, ,
\end{align}
\end{subequations}
one finds
\begin{subequations}
\label{transforqptilde2}
\begin{align}
\beta_{k}(\varphi) &= \beta_{k} \, , \label{transforqptilde2A}\\
\beta^{\dagger}_{k}(\varphi) &= \beta^{\dagger}_{k} + \sum_{k'} Z^{20}_{kk'}(\varphi) \beta_{k'}\, , \label{transforqptilde2B}
\end{align}
\end{subequations}
which can be compacted under the form of a non-unitary linear gauge-dependent Bogoliubov transformation
\begin{equation}
\left(
\begin{array} {c}
\beta(\varphi) \\
\beta^{\dagger}(\varphi)
\end{array}
\right) 
= 
\left(
\begin{array} {cc}
1 & 0 \\
Z^{20}(\varphi) &  1
\end{array}
\right) 
\left(
\begin{array} {c}
\beta \\
\beta^{\dagger}
\end{array}
\right) \, . \label{transforqptildematrix}
\end{equation}
One notes that a pure de-excitation operator is invariant under such a transformation, i.e.
\begin{align}
e^{Z(\varphi)} {\cal B}_{\mu} e^{-Z(\varphi)} = {\cal B}_{\mu} \, .
\end{align}
Fermionic anti-commutation rules of the transformed quasiparticle operators are maintained given that
\begin{subequations}
\begin{align}
 \left\{ \beta_{k}(\varphi)  , \beta_{k'}(\varphi) \right\} &= \left\{ \beta_{k}  , \beta_{k'} \right\} \nonumber\\
 &=0 \, , \\
 \left\{ \beta_{k}(\varphi)  , \beta^{\dagger}_{k'}(\varphi) \right\} &=  \left\{ \beta_{k}  , \beta^{\dagger}_{k'} \right\} + \sum_{k_1}Z^{20}_{k'k_1}(\varphi)\left\{ \beta_{k}  , \beta_{k_1} \right\} \nonumber \\
 &= \delta_{kk'}\, , \\
\left\{ \beta^{\dagger}_{k}(\varphi)  , \beta^{\dagger}_{k'}(\varphi) \right\}     &= \sum_{k_1}Z^{20}_{k'k_1}(\varphi)\left\{ \beta^{\dagger}_{k}  , \beta_{k_1} \right\} \nonumber \\
 &+ \sum_{k_2}Z^{20}_{kk_2}(\varphi)\left\{ \beta_{k_2}  , \beta^{\dagger}_{k'} \right\} \nonumber \\
&= Z^{20}_{k'k}(\varphi) + Z^{20}_{kk'}(\varphi) \nonumber \\
&= 0\, .
\end{align}
\end{subequations}

Given a normal-ordered operator defined through
\begin{equation}
W^{ij} \equiv \frac{1}{i!}\frac{1}{j!} \!\!\sum_{\substack{k_1\ldots k_i \\ l_{1}\ldots l_{j}}}\!\! W^{ij}_{k_1\ldots k_i l_{1}\ldots l_{j}} \beta^\dagger_{k_1} \ldots \beta^\dagger_{k_i}\beta_{l_{j}} \ldots \beta_{l_{1}}\, , \label{op}
\end{equation}
its similarity-transformed partner reads as\footnote{It is important to understand the notation used in the present document, i.e. $W^{(ij)}_{Z}(\varphi)$ denotes the transformed operator of $W^{ij}$ such that the notation $(ij)$ is a sole reminder of the normal-ordered nature of the original operator but does {\it not} characterize the normal-ordered nature of the transformed operator. On the other hand, $W^{mn}_{Z}(\varphi)$ does denote the normal-ordered part of the transformed operator $W_{Z}(\varphi)$ containing $m$ ($n$) quasiparticle creation (anihilation) operators. Consistently, one may use the notation $W^{mn(ij)}_{Z}(\varphi)$ to further represent the part of $W^{mn}_Z(\varphi)$ originating from $W^{ij}$; see Ref~\cite{Duguet:2015yle} for more details.}
\begin{widetext}
\begin{align}
W^{(ij)}_{Z}(\varphi) &\equiv e^{Z(\varphi)} W^{ij} e^{-Z(\varphi)} \label{rotatedop} \\
&= \frac{1}{i!}\frac{1}{j!} \!\!\sum_{\substack{k_1\ldots k_i \\ l_{1}\ldots l_{j}}}\!\! W^{ij}_{k_1\ldots k_i l_{1}\ldots l_{j}} \beta^\dagger_{k_1}(\varphi) \ldots \beta^\dagger_{k_i}(\varphi)\beta_{l_{j}}(\varphi) \ldots \beta_{l_{1}}(\varphi) \nonumber \\
&\equiv \sum_{n=j}^{i+j} \!\!\! \sum_{\substack{m=0\\m+n\leq i+j}}^{i} \!\!\! \!\!\! \frac{1}{m!}\frac{1}{n!} \!\sum_{\substack{k_1\ldots k_m \\ l_{1}\ldots l_{n}}}\!\! W^{mn(ij)}_{k_1\ldots k_m l_{1}\ldots l_{n}}(\varphi) \beta^\dagger_{k_1} \ldots \beta^\dagger_{k_m}\beta_{l_{n}} \ldots \beta_{l_{1}} \, . \nonumber
\end{align}
\end{widetext}
The last line defines a sum of normal-ordered terms obtained by applying Wick's theorem with respect to $| \Phi \rangle$ to the transformed operator. Given the nature of the transformation defined in Eq.~\ref{transforqptildematrix}, the resulting terms have at least as many annihilation operators as the original operator ($j$) and possibly up to the total number of original quasiparticle operators ($i+j$). The number of creation operators ranges from 0 to the original one ($i$) such that the overall number of quasiparticle operators is bound to remain between $j$ and $i+j$ in each term.

\section{Angle-dependent BCC equations}
\label{angledependentequations}

Let us derive the first set of equations providing access to all amplitudes but $W_0(\varphi)$, i.e. to all the ${\cal T}_n(\varphi)$. One starts from Eq.~\ref{equatmotionenergy1} obtained by left-multiplying the Schr\"odinger equation with n-tuply excited rotated Bogoliubov states and written in a compact form as
\begin{align}
\tilde{{\cal H}}_{\mu}(\varphi) &=  E^{\text{A}} \, \tilde{{\cal N}}_{\mu}(\varphi) \, . \label{angledepeqset2}
\end{align} 

\subsection{Unexcited equation}

Equation~\ref{angledepeqset2} is obviously valid for ${\cal B}_{\mu}=1$. Inserting Eq.~\ref{reducedkernels2} into it, one obtains the {\it connected} form of the energy eigenequation
\begin{align}
E^{\text{A}} &=   \langle \Phi |  H_{Z}(\varphi) e^{{\cal T}(\varphi)} | \Phi \rangle_{C}   \label{angledepeqset3} \, ,
\end{align}
where the connected Hamiltonian kernel, denoted as $h(\varphi)$ in Eq.~\ref{reducedkernels2}, naturally terminates such that the above equation formally reads as the standard, i.e., diagonal or unrotated, BCC energy equation~\cite{Signoracci:2014dia}. These connected residuals can be trivially expressed in terms of the double similarity transformed Hamiltonian
\begin{equation}
\bar{H}_Z(\varphi) \equiv e^{-\mathcal{T}(\varphi)} H_Z(\varphi)e^{+\mathcal{T}(\varphi)} \, ,
\end{equation}
according to
\begin{align}
E^{\text{A}} &=   \langle \Phi |  \bar{H}_Z(\varphi) | \Phi \rangle   \label{angledepeqset3last}  \, .
\end{align}

\subsection{Singly-excited equation}

Following the same steps as in Sec.~\ref{energy}, singly-excited kernels introduced in Eq.~\ref{excitkernelsB} can be rewritten as
\begin{subequations}
\label{rewriteexcitkernelsB}
\begin{align}
\tilde{{\cal N}}_{k_1 k_2}(\varphi)  &=  n_{k_1 k_2}(\varphi){\cal N}(\varphi) \, , \label{rewriteexcitkernelsB1} \\
\tilde{{\cal H}}_{k_1 k_2}(\varphi) &= h_{k_1 k_2}(\varphi){\cal N}(\varphi) + h(\varphi)\tilde{{\cal N}}_{k_1 k_2}(\varphi)  \, , \label{rewriteexcitkernelsB2} 
\end{align}
\end{subequations}
in terms of {\it connected} singly-excited kernels defined through
\begin{subequations}
\label{connectexcitkernelseq}
\begin{align}
n_{k_1 k_2}(\varphi)  &\equiv \langle \Phi^{k_1 k_2} |   e^{{\cal T}(\varphi)} | \Phi \rangle \, , \label{connectexcitkernelseqA} \\
h_{k_1 k_2}(\varphi) &\equiv \langle \Phi^{k_1 k_2} | H_{Z}(\varphi)  e^{{\cal T}(\varphi)} | \Phi \rangle_{C} \label{connectexcitkernelseqB} \\
&= \langle \Phi^{k_1 k_2} | \bar{H}_Z(\varphi) | \Phi \rangle \, ,  \nonumber 
\end{align}
\end{subequations}
where the connected character stipulates that the cluster operator is necessarily contracted with the operator $H_{Z}(\varphi)$.

Inserting Eq.~\ref{rewriteexcitkernelsB} into  Eq.~\ref{angledepeqset2} and exploiting Eq.~\ref{angledepeqset3}, one obtains the {\it connected} single equation under the form
\begin{align}
0&= h_{k_1 k_2}(\varphi) \, , \label{angledepeqset4} 
\end{align}
where $h_{k_1 k_2}(\varphi)$ naturally terminates such that Eq.~\ref{angledepeqset4} formally reads exactly as the standard, i.e., diagonal or unrotated, single BCC amplitude equation~\cite{Signoracci:2014dia}. 

\subsection{Doubly-excited equation}

To unreveil the hierarchy of angle-dependent BCC equations, the doubly-excited one is now worked out. Following the same steps as before, doubly-excited kernels introduced in Eq.~\ref{excitkernelsB} can be rewritten as
\begin{subequations}
\label{rewriteexcitkernelsC}
\begin{align}
\tilde{{\cal N}}_{k_1 k_2k_3 k_4}(\varphi)  &=  n_{k_1 k_2k_3 k_4}(\varphi){\cal N}(\varphi) \, , \label{rewriteexcitkernelsC1} \\
\tilde{{\cal H}}_{k_1 k_2k_3 k_4}(\varphi) &= h_{k_1 k_2k_3 k_4}(\varphi){\cal N}(\varphi) \nonumber \\
& + h_{k_1 k_2}(\varphi)  \tilde{{\cal N}}_{k_3 k_4}(\varphi) \nonumber \\
& - h_{k_1 k_3}(\varphi)  \tilde{{\cal N}}_{k_2 k_4}(\varphi) \nonumber \\
& + h_{k_1 k_4}(\varphi)  \tilde{{\cal N}}_{k_2 k_3}(\varphi) \nonumber \\
& + h_{k_2 k_3}(\varphi)  \tilde{{\cal N}}_{k_1 k_4}(\varphi) \nonumber \\
& - h_{k_2 k_4}(\varphi)  \tilde{{\cal N}}_{k_1 k_3}(\varphi) \nonumber \\
& + h_{k_3 k_4}(\varphi)  \tilde{{\cal N}}_{k_1 k_2}(\varphi) \nonumber \\
& + h(\varphi)\tilde{{\cal N}}_{k_1 k_2k_3 k_4}(\varphi)  \, , \label{rewriteexcitkernelsC2} 
\end{align}
\end{subequations}
where {\it connected} doubly-excited kernels are defined as
\begin{subequations}
\label{connectexcitkernelseq2}
\begin{align}
n_{k_1 k_2k_3 k_4}(\varphi)  &\equiv \langle \Phi^{k_1 k_2k_3 k_4} |   e^{{\cal T}(\varphi)} | \Phi \rangle \, , \label{connectexcitkernelseqA2} \\
h_{k_1 k_2k_3 k_4}(\varphi) &\equiv \langle \Phi^{k_1 k_2k_3 k_4} | H_{Z}(\varphi)  e^{{\cal T}(\varphi)} | \Phi \rangle_{C} \label{connectexcitkernelseqB2} \\
&= \langle \Phi^{k_1 k_2k_3 k_4} | \bar{H}_Z(\varphi) | \Phi \rangle \, .  \nonumber 
\end{align}
\end{subequations}

Inserting Eq.~\ref{rewriteexcitkernelsC} into  Eq.~\ref{angledepeqset2}  and exploiting Eqs.~\ref{angledepeqset3} and~\ref{angledepeqset4}, one obtains the {\it connected} double equation under the form
\begin{align}
0& = h_{k_1 k_2k_3 k_4}(\varphi) \, , \label{angledepeqset5}
\end{align}
which naturally terminates such that Eq.~\ref{angledepeqset5} formally reads exactly as the standard, i.e., diagonal or unrotated, double BCC amplitude equation~\cite{Signoracci:2014dia}. 

\subsection{$n$-tuply excited equations}

As for connected singly- and doubly-excited equations, the derivation of the $n$-tuple equation invokes all equations of lower rank. Reasoning by recurrence, one can prove that
\begin{align}
0&= h_{k_1 \ldots k_{2n}}(\varphi) \, , \label{angledepeqsetn}
\end{align}
where the naturally terminating connected n-tuply excited kernel of the Hamiltonian is defined as 
\begin{align}
h_{k_1 \ldots k_{2n}}(\varphi) &\equiv \langle \Phi^{k_1 \ldots k_{2n}} | H_{Z}(\varphi)  e^{{\cal T}(\varphi)} | \Phi \rangle_{C}  \label{connectexcitkernelseq5} \\
&= \langle \Phi^{k_1 \ldots k_{2n}} | \bar{H}_Z(\varphi) | \Phi \rangle  \, . \nonumber
\end{align}

\section{Differential equations}
\label{diffequations}

We derive now the second set of equations providing access to all angle-dependent amplitudes $W_n(\varphi)$, i.e. to $W_0(\varphi)$ and ${\cal T}_n(\varphi)$ for $n \geq 1$. 

\subsection{Master equation}

Taking the derivative of the n-tuply excited norm kernel with respect to the gauge angle gives
\begin{align}
\frac{d }{d\varphi} \tilde{{\cal N}}_{\mu}(\varphi) &= \frac{d }{d\varphi} \langle \Phi |  R(\varphi) {\cal B}_{\mu} \, e^{U} | \Phi \rangle = i \breve{{\cal A}}_{\mu}(\varphi) \label{physics3bis}  \, .
\end{align}

\subsection{Particle number kernel}

The particle number operator is defined in an arbitrary single-particle basis through
\begin{equation}
A\equiv \sum_{pq}  a^{11}_{pq} c^{\dagger}_p c_q \, ,
\end{equation}
with $a^{11}_{pq} = \delta_{pq} $. In the quasiparticle basis of $|  \Phi \rangle $, the normal-ordered form of $A$ reads as
\begin{align}
 A &\equiv A^{00} + A^{11} + A^{20} + A^{02} \label{operatorAinqp} \\
&\equiv A^{00} + \sum_{k_1 k_2}  A^{11}_{k_1 k_2} \beta^{\dagger}_{k_1} \beta_{k_2} \nonumber \\
   & \,\,\,\,\,\, + \frac{1}{2} \sum_{k_1 k_2}  \left\{A^{20}_{k_1 k_2} \beta^{\dagger}_{k_1} \beta^{\dagger}_{k_2} 
                                                                                              + A^{02}_{k_1 k_2} \beta_{k_2} \beta_{k_1} \right\} \, ,\nonumber 
\end{align}
where the matrix elements of the various normal-ordered parts are given in terms of the Bogoliubov transformation defining $|  \Phi \rangle $ via
\begin{subequations}
\label{transfomatricesAsymgaugecorps}
\begin{align}
A^{00} &= \text{Tr}(V^{\ast}V^T) \, , \label{transfomatricesAsymgaugecorps1} \\
A^{11}_{k_1 k_2} &= \left[U^\dagger U - V^\dagger V\right]_{k_1 k_2} \, , \label{transfomatricesAsymgaugecorps2} \\
A^{20}_{k_1 k_2} &= \left[U^\dagger V^* - V^\dagger U^*\right]_{k_1 k_2} \, , \label{transfomatricesAsymgaugecorps3} \\
A^{02}_{k_1 k_2} &= \left[U^T  V - V^T  U \right]_{k_1 k_2}  \, . \label{transfomatricesAsymgaugecorps4}
\end{align}
\end{subequations}
In agreement with Eq.~\ref{rotatedop}, the similarity-transformed particle-number operator is written as
\begin{align}
 A_{Z}(\varphi) &\equiv A^{00}_{Z}(\varphi) + A^{11}_{Z}(\varphi) + A^{20}_{Z}(\varphi) + A^{02}_{Z}(\varphi) \label{operatorAinqptransfo} \\
  &\equiv A^{00}(\varphi) + \sum_{k_1 k_2}  A^{11}_{k_1 k_2}(\varphi) \beta^{\dagger}_{k_1} \beta_{k_2} \nonumber \\
   & \,\,\,\,\,\, + \frac{1}{2} \sum_{k_1 k_2}  \left\{A^{20}_{k_1 k_2}(\varphi) \beta^{\dagger}_{k_1} \beta^{\dagger}_{k_2} 
                                                                                              + A^{02}_{k_1 k_2}(\varphi) \beta_{k_2} \beta_{k_1} \right\} \, , \nonumber 
\end{align}
where the expression of each angle-dependent matrix element can be found in Ref~\cite{Duguet:2015yle}.

With the above definitions at hand, the particle-number kernel introduced in Eq.~\ref{newkernelsA} can be re-expressed as
\begin{align}
\breve{{\cal A}}_{\mu}(\varphi) 
&= \langle \Phi(\varphi) | \Phi \rangle \langle \Phi |  A_{Z}(\varphi) {\cal B}_{\mu} \, e^{U_{Z}(\varphi)} | \Phi \rangle \nonumber \\
&= A^{00}_{Z}(\varphi) \tilde{{\cal N}}_{\mu}(\varphi) + \langle \Phi(\varphi) | \Phi \rangle \langle \Phi |  A^{02}_{Z}(\varphi) {\cal B}_{\mu} \, e^{U_{Z}(\varphi)} | \Phi \rangle \nonumber \\
&= A^{00}_{Z}(\varphi) \tilde{{\cal N}}_{\mu}(\varphi) + \langle \Phi(\varphi) | \Phi \rangle \langle \Phi |   {\cal B}_{\mu} A^{02}_{Z}(\varphi) \, e^{U_{Z}(\varphi)} | \Phi \rangle \nonumber \\
&= A^{00}_{Z}(\varphi) \tilde{{\cal N}}_{\mu}(\varphi) + {\cal N}(\varphi) \langle \Phi^{\mu} |  A^{02}_{Z}(\varphi) \, e^{{\cal T}(\varphi)} | \Phi \rangle \nonumber \\
&= A^{00}_{Z}(\varphi) \tilde{{\cal N}}_{\mu}(\varphi) + \tilde{{\cal A}}^{02}_{\mu}(\varphi)    \, , \label{relationkernels}
\end{align}
where the kernel $\tilde{{\cal A}}^{02}_{\mu}(\varphi)$ is defined similarly to the Hamiltonian kernel introduced in Eq.~\ref{excitkernelsB}.

\subsection{Unexcited amplitude equation}

Applying Eqs.~\ref{physics3bis}-\ref{relationkernels} for ${\cal B}_{\mu}=1$, the unexcited norm kernel satisfies
\begin{align}
\frac{d }{d\varphi} {\cal N}(\varphi) &=  i\big(A^{00}(\varphi) {\cal N}(\varphi) + \tilde{{\cal A}}^{02}(\varphi)\big)  \, . \label{diffequationskernels1} 
\end{align}
Introducing the connected unexcited kernel of $A^{02}$ 
\begin{align}
a^{02}(\varphi) &\equiv \langle \Phi |  A^{02}_{Z}(\varphi) e^{{\cal T}(\varphi)} | \Phi \rangle_{C} \nonumber \\
&=  \frac{1}{2} \sum_{k_1 k_2}  A^{02}_{k_1 k_2}(\varphi)W_{k_1 k_2}(\varphi)\, , \label{linkedkernelA2} 
\end{align}
and further noticing that
\begin{align}
a(\varphi) &= A^{00}(\varphi)+a^{02}(\varphi) \, , 
\end{align}
Eq.~\ref{diffequationskernels1} can be rewritten as a differential equation for the unexcited norm kernel 
\begin{align}
\frac{d }{d\varphi} {\cal N}(\varphi) &=  i \, a(\varphi) {\cal N}(\varphi) \, . \label{diffequationskernels1bis} 
\end{align}
Given that ${\cal N}(\varphi)=\langle \Phi (\varphi)| \Phi \rangle \, n(\varphi)$ and that
\begin{align}
\frac{d }{d\varphi} \langle \Phi (\varphi)| \Phi \rangle &= i \langle \Phi (\varphi)| A | \Phi \rangle  \nonumber \\
&= i \langle \Phi (\varphi)| \Phi \rangle \langle \Phi | A_{Z}(\varphi) | \Phi \rangle  \nonumber \\
&= i A^{00}(\varphi)\langle \Phi (\varphi)| \Phi \rangle  \, . \label{uncorrelatednorm} 
\end{align}
one obtains the differential equation for the correlated part of the norm kernel
\begin{align}
\frac{d }{d\varphi} n(\varphi) &= i \, a^{02}(\varphi)n(\varphi)  \, . \label{diffequationskernels2bis} 
\end{align}
Exploiting that $n(\varphi)=e^{W_0(\varphi)}$, $W_{0}(\varphi)$ is eventually shown to satisfy the first-order differential equation
\begin{align}
\frac{d }{d\varphi} W_0(\varphi) &=  i \, a^{02}(\varphi)  \, , \label{diffequationskernels2ter} 
\end{align}
which connects it to $W_{1}(\varphi)$ through Eq.~\ref{linkedkernelA2}.

\subsection{Singly-excited amplitude equation}

Applying Eqs.~\ref{physics3bis}-\ref{relationkernels} for ${\cal B}_{\mu}={\cal B}_{k_1 k_2}$, the singly-excited norm kernel satisfies
\begin{align}
\frac{d }{d\varphi} \tilde{{\cal N}}_{k_1 k_2}(\varphi) &=  -i\big(A^{00}(\varphi) \tilde{{\cal N}}_{k_1 k_2}(\varphi) + \tilde{{\cal A}}^{02}_{k_1 k_2}(\varphi)\big)  \label{diffequationskernels3} \\
&=  -i\big(a(\varphi)\tilde{{\cal N}}_{k_1 k_2}(\varphi) +  a^{02}_{k_1 k_2}(\varphi) {\cal N}(\varphi)\big) \, , \nonumber
\end{align}
where the connected singly-excited kernel of $A^{02}$ reads as
\begin{align}
a^{02}_{k_1 k_2}(\varphi) &\equiv \langle \Phi^{k_1 k_2} |  A^{02}_{Z}(\varphi) e^{{\cal T}(\varphi)} | \Phi \rangle_{C} \label{singlylinkedkernelA2}  \\
&=  \sum_{k_3 k_4} A^{02}_{k_3 k_4}(\varphi) \Big[ \frac{1}{2}W_{k_3 k_4k_1 k_2 }(\varphi) \nonumber \\
&\hspace{2.5cm} - W_{k_1 k_3}(\varphi)W_{k_2 k_4}(\varphi) \Big]  \, , \nonumber
\end{align}
and where it was exploited that
\begin{align}
\tilde{{\cal A}}^{02}_{k_1 k_2}(\varphi) &\equiv  a^{02}_{k_1 k_2}(\varphi){\cal N}(\varphi)  + a^{02}(\varphi)\tilde{{\cal N}}_{k_1 k_2}(\varphi)\, . \label{exploit}
\end{align}
Using Eqs.~\ref{rewriteexcitkernelsB1} and~\ref{diffequationskernels1bis}, along with the fact that
\begin{align}
n_{k_1 k_2}(\varphi) &=  W_{k_1 k_2}(\varphi)   \, , \label{diffequationskernels4}
\end{align}
$W_{1}(\varphi)$ is eventually shown to satisfy the first-order differential equation
\begin{align}
\frac{d }{d\varphi} W_{k_1 k_2}(\varphi) &= i a^{02}_{k_1 k_2}(\varphi) \label{diffequationskernels5} \, ,
\end{align}
which connects it to $W_{2}(\varphi)$ through Eq.~\ref{singlylinkedkernelA2}.

\subsection{Doubly-excited amplitude equation}

Applying Eqs.~\ref{physics3bis}-\ref{relationkernels} for ${\cal B}_{\mu}={\cal B}_{k_1 k_2k_3 k_4}$, the doubly-excited norm kernel satisfies
\begin{align}
\frac{d }{d\varphi} \tilde{{\cal N}}_{k_1 k_2k_3 k_4}(\varphi) &=  i\big(A^{00}(\varphi) \tilde{{\cal N}}_{k_1 k_2k_3 k_4}(\varphi) + \tilde{{\cal A}}^{02}_{k_1 k_2k_3 k_4}(\varphi)\big)  \label{doublydiffeq1}  \, .
\end{align}
Using Eq.~\ref{rewriteexcitkernelsC1} and exploiting that
\begin{subequations}
\begin{align}
n_{k_1 k_2k_3 k_4}(\varphi) &=  W_{k_1 k_2k_3 k_4}(\varphi) \nonumber \\
& + W_{k_1 k_2}(\varphi) \, W_{k_3 k_4}(\varphi) \nonumber \\
&  - W_{k_1 k_3}(\varphi) \, W_{k_2 k_4}(\varphi) \nonumber \\
& +  W_{k_1 k_4}(\varphi) \, W_{k_2 k_3}(\varphi) \, , \label{doublydiffeq2} \\
\bar{{\cal A}}^{02}_{k_1 k_2k_3 k_4}(\varphi) &\equiv  a^{02}_{k_1 k_2k_3 k_4}(\varphi){\cal N}(\varphi) \nonumber \\
& + a^{02}_{k_1 k_2}(\varphi)  \tilde{{\cal N}}_{k_3 k_4}(\varphi) \nonumber \\
& - a^{02}_{k_1 k_3}(\varphi)  \tilde{{\cal N}}_{k_2 k_4}(\varphi) \nonumber \\
& + a^{02}_{k_1 k_4}(\varphi)  \tilde{{\cal N}}_{k_2 k_3}(\varphi) \nonumber \\
& + a^{02}_{k_2 k_3}(\varphi)  \tilde{{\cal N}}_{k_1 k_4}(\varphi) \nonumber \\
& - a^{02}_{k_2 k_4}(\varphi)  \tilde{{\cal N}}_{k_1 k_3}(\varphi) \nonumber \\
& + a^{02}_{k_3 k_4}(\varphi)  \tilde{{\cal N}}_{k_1 k_2}(\varphi) \nonumber \\
& + a^{02}(\varphi)\tilde{{\cal N}}_{k_1 k_2k_3 k_4}(\varphi)\, , \label{doublydiffeq3}
\end{align}
\end{subequations}
along with the differential equations associated with unexcited and singly-excited norm kernels derived above, $W_{2}(\varphi)$ is shown to satisfy the first-order differential equation
\begin{align}
\frac{d }{d\varphi} W_{k_1 k_2k_3 k_4}(\varphi) &= i a^{02}_{k_1 k_2k_3 k_4}(\varphi) \, , \label{doublydiffeq4}
\end{align}
where the connected doubly-excited kernel of $A^{02}$
\begin{align}
a^{02}_{k_1 k_2k_3 k_4}(\varphi) &\equiv \langle \Phi^{k_1 k_2k_3 k_4} |  A^{02}_Z(\varphi) e^{{\cal T}(\varphi)} | \Phi \rangle_{C} \label{doublydiffeq5}  \\
&=   \sum_{k_5 k_6} A^{02}_{k_5 k_6}(\varphi) \big[ \frac{1}{2}W_{k_5k_6k_1  k_2 k_3 k_4}(\varphi)\nonumber \\
&\hspace{2.5cm} +W_{k_1 k_5}(\varphi) W_{k_6 k_2 k_3 k_4}(\varphi) \nonumber \\
&\hspace{2.5cm} + W_{k_2 k_5}(\varphi) W_{k_1 k_6 k_3 k_4}(\varphi) \nonumber \\
&\hspace{2.5cm}  + W_{k_3 k_5}(\varphi) W_{k_1 k_2 k_6 k_4}(\varphi) \nonumber \\
&\hspace{2.5cm}  + W_{k_4 k_5}(\varphi) W_{k_1 k_2 k_3 k_6}(\varphi) \big]  \, , \nonumber
\end{align}
connects it to $W_{3}(\varphi)$.

\subsection{$n$-tuple amplitude equation}

As for single and double amplitudes, the derivation of the differential equation satisfied by the $n$-tuple amplitude invokes all those lower ranks. Reasoning by recurrence, one can prove
\begin{align}
\frac{d }{d\varphi} W_{k_1 \ldots k_{2n}}(\varphi) &= i a^{02}_{k_1 \ldots k_{2n}}(\varphi) \, , \label{ntuplydiffeq1}
\end{align}
where the connected $n$-tuply excited kernel of $A^{02}$
\begin{align}
a^{02}_{k_1 \ldots k_{2n}}(\varphi) &\equiv \langle \Phi^{k_1 \ldots k_{2n}} |  A^{02}_Z(\varphi) e^{{\cal T}(\varphi)} | \Phi \rangle_{C} \label{ntuplydiffeq2}    \, ,
\end{align}
connects it to $W_{n+1}(\varphi)$.

\subsection{Initial conditions}

The ODEs need to be complemented with initial conditions. Working up to the n-tuple level, one most naturally employs the initial conditions
\begin{subequations}
\label{initialconditionsA} 
\begin{align}
W_0(0) &= 0 \label{initialconditionsA1}    \, , \\
W_{k}(0) &= U_{k} \label{initialconditionsA2}    \, ,
\end{align}
\end{subequations}
for $k\leq n$ and $W_{k}(0)=0$ for $k>n$, where the gauge-unrotated amplitudes $U_k$ are to be determined by solving standard BCC equations (Eq.~\ref{angleindepBCCequat}) at the same n-tuple level.

\section{Off-diagonal kernels}
\label{AppEthan}

\subsection{Results}

\begin{figure*}[t]
\includegraphics[width=0.96\columnwidth]{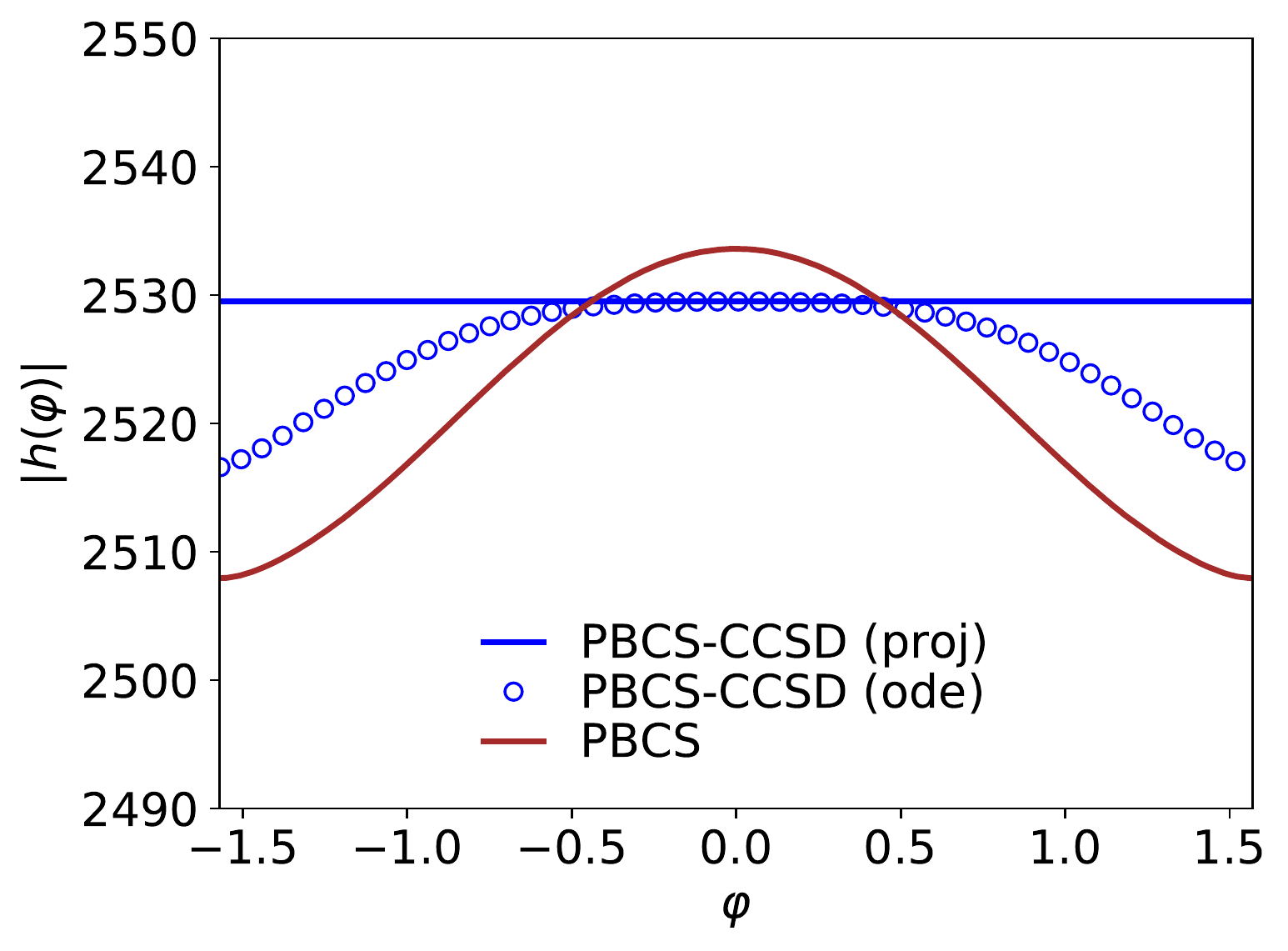}
\hfill
\includegraphics[width=0.96\columnwidth]{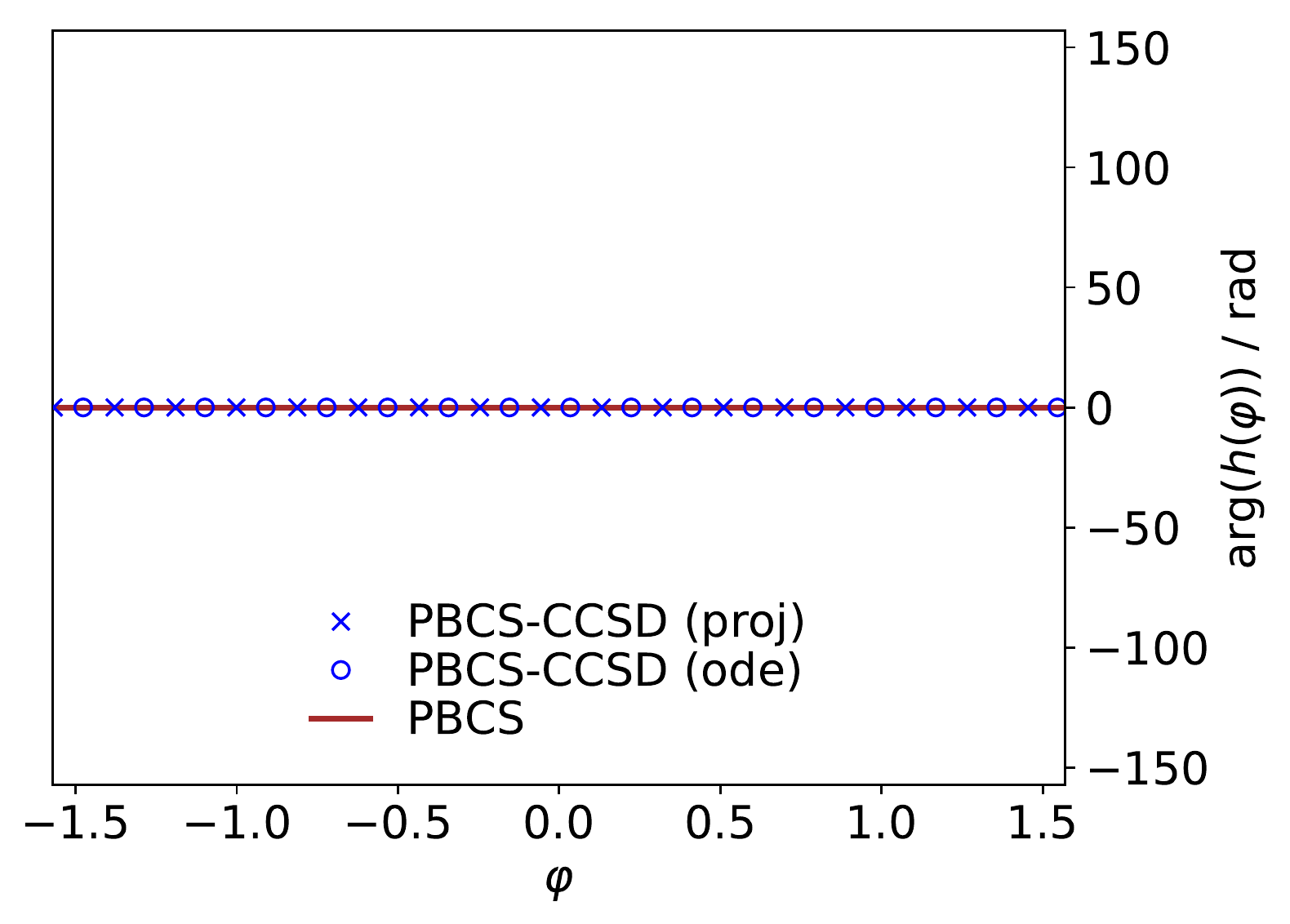}
\caption{(Color online) Connected part $h(\varphi)$ of the off-diagonal hamiltonian kernel as a function of the gauge angle. Left panel: norm of $h(\varphi)$ in units of $\Delta \epsilon$. Right panel: phase of $h(\varphi)$ in radians. The calculation is performed for 100 levels at half filling using $G = 1.5G_c = 0.27336$. Kernels are displayed at the PBCS and PBCS-CCSD levels. At the PBCS-CCSD level, gauge-dependent cluster amplitudes are either obtained from the gauge-dependent BCC equations ("proj") introduced in Sec.~\ref{ampli1} or from the ODEs ("ode") introduced in Sec.~\ref{ampli2}.
\label{Fig:hkernel}}
\end{figure*}

\begin{figure*}
\includegraphics[width=0.96\columnwidth]{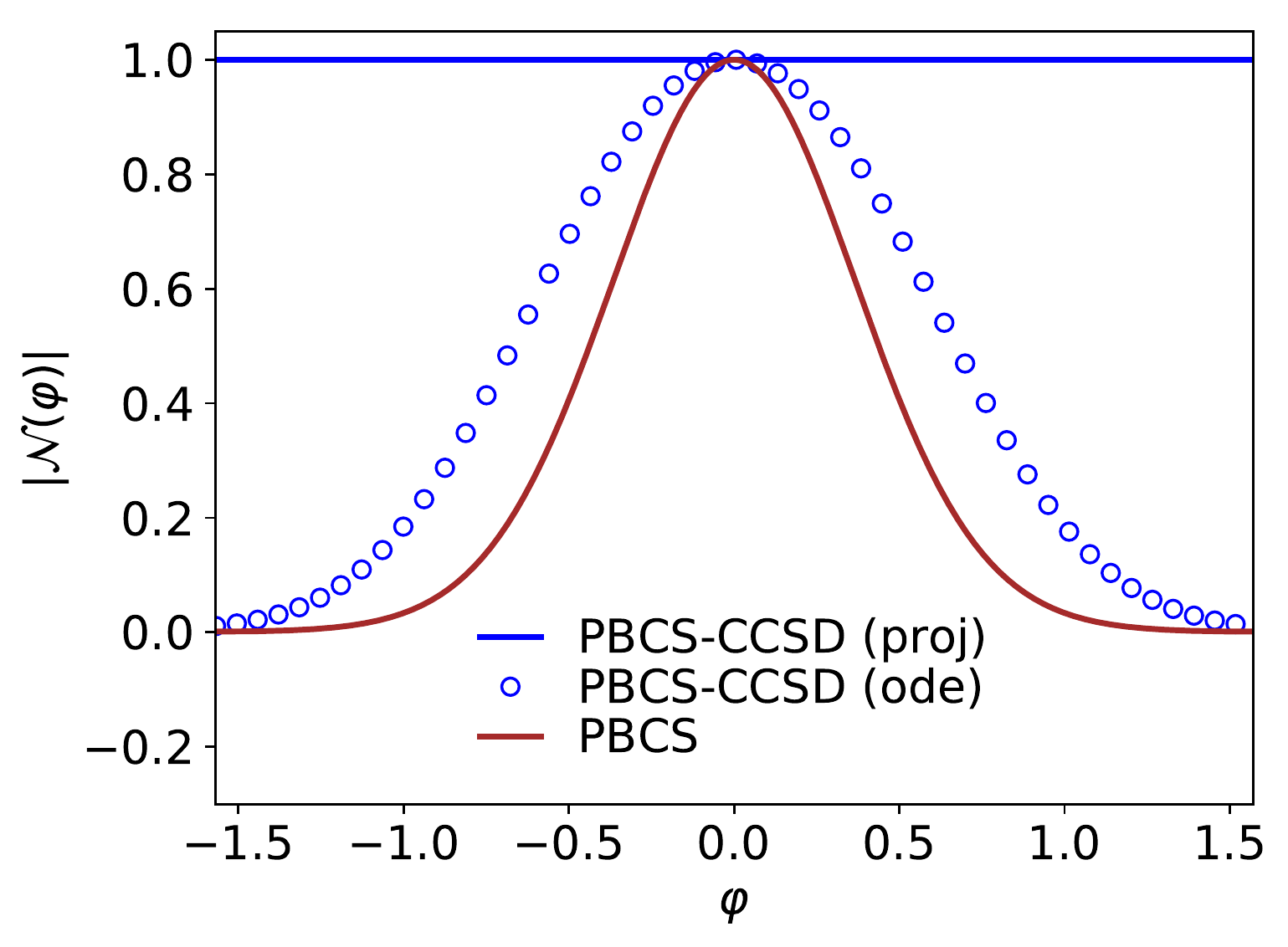}
\hfill
\includegraphics[width=0.96\columnwidth]{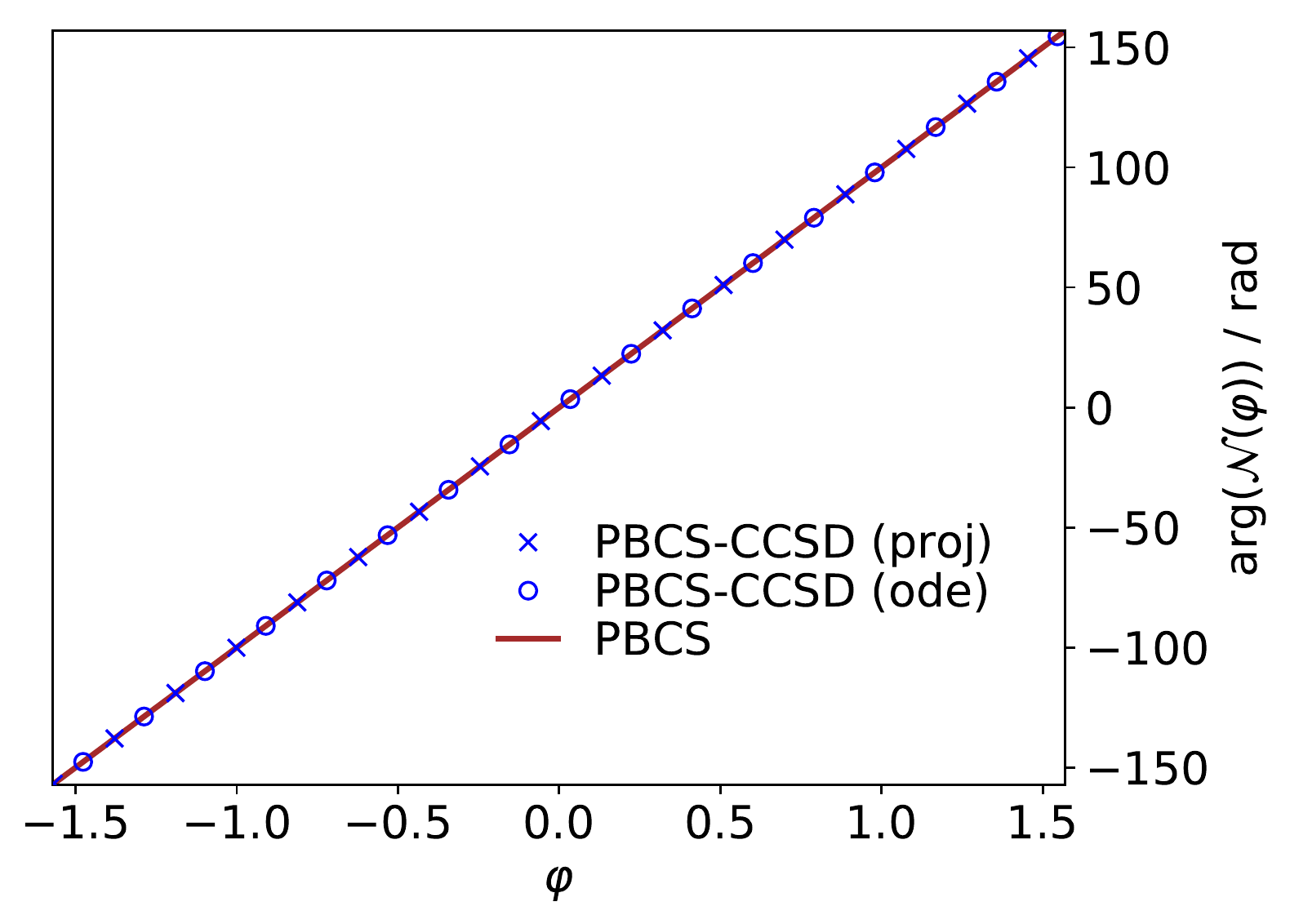}
\caption{(Color online) Same as Fig.~\ref{Fig:hkernel} for the norm kernel ${\cal N}(\varphi)$.
\label{Fig:nkernel}}
\end{figure*}

The connected part $h(\varphi)$ of the off-diagonal hamiltonian kernel and the norm kernel ${\cal N}(\varphi)$ are displayed as a function of the gauge angle in Figs.~\ref{Fig:hkernel} and~\ref{Fig:nkernel}, respectively. Results are generated for the pairing Hamiltonian with 100 levels at half filling ($G/G_c = 1.5$) at the PBCS and PBCS-CCSD levels. 

Before discussing the results, it is worth noting that in an exact\footnote{This is more generally true in any symmetry-conserving limit of PBCC.} limit of PBCC, the connected part of the off-diagonal hamiltonian or particle number kernel is independent of the gauge angle~\cite{Duguet:2014jja,Duguet:2015yle} while the norm kernel is nothing but ${\cal N}(\varphi)=e^{iA\varphi}$. Contrarily, the non-trivial impact of the particle-number projection precisely relates to the gauge-angle dependency acquired by $h(\varphi)$ and $a(\varphi)$ as well as to the departure of ${\cal N}(\varphi)$ from a single IRREP of the $U(1)$ group as a result of the underlying symmetry breaking. 

The results obtained for at the well-celebrated PBCS ("MF") level provides a reference. The norm of $h(\varphi)$ displays a typical bell-shaped  curve that is maximum at $\varphi=0$\footnote{Note that $h(0)$ is real and delivers nothing but the total unprojected BCS energy.}. The norm kernel displays a similar qualitative behavior with ${\cal N}(0)=1$ as a result of $\langle \Phi | \Phi \rangle$. The phase of both $h(\varphi)$ and ${\cal N}(\varphi)$ are strictly linear in $\varphi$ but while it is zero for $h(\varphi)$ that is real, the slope is equal to A for ${\cal N}(\varphi)$.

Moving now to PBCS-CCSD, let us first focus on the case where ${\cal T}_1(\varphi)$ and  ${\cal T}_2(\varphi)$, from which $h(\varphi)$ is computed according to Eq.~\ref{reducedkernels2}, are obtained by solving the ODEs introduced in Sec.~\ref{ampli2}. The behavior of $h(\varphi)$ is qualitatively very similar to what is observed at the PBCS level, i.e. when $U=W(\varphi)= 0$. While the phase remains strictly null, the norm $h(\varphi)$ is more spread out with a maximum still located at $\varphi=0$\footnote{Note that $h(0)$ is real and delivers nothing but the total unprojected BCS-CCSD energy.}. By going from PBCS to PBCS-CCSD, one starts observing the flattening of $h(\varphi)$ and the converging towards the exact solution that must occur as one includes higher-order cluster operators\footnote{We recall that in the exact limit $h(\varphi)$ is flat and equal to the actual ground-state energy.}. As for $h(\varphi)$, the norm of ${\cal N}(\varphi)$ resembles the one at the PBCS level but is flatter as a result of included many-body correlations\footnote{We recall that $|{\cal N}(\varphi)|=1$ in the exact limit.}. The fact that ${\cal N}(0)=1$ remains true testifies of the intermediate normalization satisfied by the BCC wave function at any truncation order.

Let us now focus on the case where ${\cal T}_1(\varphi)$ and ${\cal T}_2(\varphi)$ are obtained by solving the gauge-dependent BCC amplitude equations introduced in Sec.~\ref{ampli1}. One observes that $h(\varphi)$ is constant and real\footnote{The same result holds for the connected part of the particle number kernel, i.e. $a(\varphi)=a(0)$ for all $\varphi$, or to any scalar operator under gauge transformations. See Sec.~\ref{proof} below.}, i.e.  $h(\varphi)=h(0)$ for all $\varphi$. This behavior is thus entirely different from what is obtained at the lowest order, i.e. PBCS level, which is both surprising and at odd with the results obtained from the ODEs. An analytical proof of this unexpected result is given in Sec.~\ref{proof} below. By virtue of Eq.~\ref{diffequationskernels1bis}, the norm kernel behaves trivially as ${\cal N}(\varphi)=e^{ia(0)\varphi}=e^{iA\varphi}$ is this case. As mentioned above, the actual independence of $h(\varphi)$ on the gauge angle renders the particle number projection trivial, i.e. inactive, even though the cluster operator $U$ is truncated such that the symmetry is indeed broken at the BCC level. While requiring that ${\cal T}(\varphi)$ satisfies angle-dependent BCC equations is correct in the exact limit, it is not pertinent as soon as $U$ is truncated. Indeed, this demand happens to break completely the link to the original definition of ${\cal T}(\varphi)$ in Eq.~\ref{angledependentclusters} in which its dependence on the gauge angle is driven by a gauge rotation. In conclusion, the determination of ${\cal T}(\varphi)$ via angle-dependent BCC amplitude equations is inappropriate and must be discarded in actual calculations where the symmetry is broken and needs to be effectively restored. 

\subsection{Constant connected hamiltonian kernel}
\label{proof}

Using the disentanglement properties of the Lie algebra~\cite{Gilmore2}, the symmetry rotation operator $R(\varphi)$ can always be decomposed exactly as
\begin{equation}
R(\varphi) = e^{X(\varphi)} \, e^{Y(\varphi)} \, e^{Z(\varphi)}
\end{equation}
where
\begin{subequations}
\begin{align}
X(\varphi) &= \frac{1}{2} \, \sum_{k_1 k_2} X_{k_1 k_2}(\varphi) \, \beta_{k_1}^\dagger \, \beta_{k_2}^\dagger,
\\
Y(\varphi) &= Y_0(\varphi) + \sum_{k_1 k_2} Y_{k_1 k_2}(\varphi) \, \beta_{k_1}^\dagger \, \beta_{k_2},
\\
Z(\varphi) &= \frac{1}{2} \, \sum_{k_1 k_2} Z_{k_1 k_2}(\varphi) \, \beta_{k_1} \, \beta_{k_2}.
\end{align}
\end{subequations}
In other words, $Z(\varphi)$ destroys a pair of quasiparticles, $X(\varphi)$ creates a pair of quasiparticles, and $Y(\varphi)$ preserves the number of quasiparticles.  The precise details of the matrix elements need not concern us here, but note that $Z(\varphi)$ is nothing but the Thouless operator introduced in App.~\ref{SecThouless}.

Because the Hamiltonian is invariant under gauge rotation, so that
\begin{equation}
R(\varphi) \, H \, R^{-1}(\varphi) = H,
\end{equation}
the Thouless-transformed Hamiltonian $H_Z(\varphi)$ can be written as
\begin{equation}
H_Z(\varphi) = e^{Z(\varphi)} \, H \, e^{-Z(\varphi)}
 = e^{-Y(\varphi)} \, e^{-X(\varphi)} \, H \, e^{X(\varphi)} \, e^{Y(\varphi)}.
\label{Eq:Hz}
\end{equation}

We can use this complete disentanglement to write alternative forms for the norm and Hamiltonian kernels given in Eq. \ref{kernelsA}.  The norm kernel is
\begin{subequations}
\begin{align}
\mathcal{N}(\varphi)
 &= \langle \Phi| R(\varphi) \, e^U |\Phi\rangle
\\
 &= \langle \Phi| e^{X(\varphi)} \, e^{Y(\varphi)} \, e^{Z(\varphi)} \, e^U |\Phi\rangle
\\
 &= e^{Y_0(\varphi) + W_0(\varphi)} \, \langle \Phi| e^{\mathcal{T}(\varphi)} |\Phi\rangle
\\
 &= e^{Y_0(\varphi) + W_0(\varphi)}
\label{Eq:NTransformed}
\end{align}
\end{subequations}
while the Hamiltonian kernel becomes
\begin{subequations}
\begin{align}
\mathcal{H}(\varphi)
 &= \langle \Phi| R(\varphi) \, H \, e^U |\Phi\rangle
\\
 &= \langle \Phi| H \, R(\varphi) \, e^U |\Phi\rangle
\\
 &= \langle \Phi| H \, e^{X(\varphi)} \, e^{Y(\varphi)} \, e^{Z(\varphi)} \, e^U |\Phi\rangle
\\
 &= \langle \Phi| H \, e^{X(\varphi)} \, e^{Y(\varphi)} \, e^{\mathcal{T}(\varphi)} |\Phi\rangle \, e^{W_0(\varphi)}.
\end{align}
\end{subequations}

To simplify the Hamiltonian kernel, we can define
\begin{equation}
\tilde{\mathcal{T}}(\varphi) = e^{Y(\varphi)} \, \mathcal{T}(\varphi) \, e^{-Y(\varphi)}.
\end{equation}
Because $\tilde{\mathcal{T}}(\varphi)$ contains only quasiparticle creation operators, it commutes with $X(\varphi)$, and it will prove convenient to also introduce
\begin{equation}
\hat{\mathcal{T}}(\varphi) = X(\varphi) + e^{Y(\varphi)} \, \mathcal{T}(\varphi) \, e^{-Y(\varphi)}.
\end{equation}
Note that $\hat{\mathcal{T}}(\varphi)$ contains all the same excitation levels as does $\mathcal{T}(\varphi)$, and also contains single excitations even if $\mathcal{T}(\varphi)$ does not, simply due to the presence of $X(\varphi)$.

Now, with these definitions in hand the Hamiltonian kernel is
\begin{subequations}
\begin{align}
\mathcal{H}(\varphi)
 &= \langle \Phi| H \, e^{X(\varphi)} \, e^{\tilde{\mathcal{T}}(\varphi)} \, e^{Y(\varphi)} |\Phi\rangle \, e^{W_0(\varphi)}
\\
 &= \langle \Phi| H \, e^{\hat{\mathcal{T}}(\varphi)} |\Phi\rangle \, e^{Y_0(\varphi) + W_0(\varphi)},
\label{Eq:HTransformed}
\end{align}
\end{subequations}
where in the second line we have noted that
\begin{equation}
e^{Y(\varphi)} |\Phi\rangle = |\Phi\rangle \, e^{Y_0(\varphi)},
\end{equation}
and the connected Hamiltonian kernel $h(\varphi)$ defined in Eq. \ref{reducedkernels2} is just
\begin{equation}
h(\varphi) = \langle \Phi| H \, e^{\hat{\mathcal{T}}(\varphi)} |\Phi\rangle.
\label{Eqn:LittleH}
\end{equation}
This identity is true regardless of the truncation of the broken symmetry cluster operator $U = \mathcal{T}(\varphi=0)$.

Thus far all we have done is to note that, regardless of the truncation of the broken symmetry cluster operator, the reduced Hamiltonian kernel can be written in terms of the \textit{untransformed} Hamiltonian and an alternative angle-dependent cluster operator $\hat{\mathcal{T}}(\varphi)$.  Our next step is to show that if $\mathcal{T}(\varphi)$ satisfies the angle-dependent BCC equations, then $\hat{\mathcal{T}}(\varphi)$ satisfies the angle-independent BCC equations, and the reduced Hamiltonian kernel is therefore constant.

To see this, suppose that $\mathcal{T}(\varphi)$ satisfies the angle-dependent BCC equations, so that
\begin{equation}
\langle \Phi^{\mu} | e^{-\mathcal{T}(\varphi)} \, H_Z(\varphi) \, e^{\mathcal{T}(\varphi)} |\Phi \rangle = 0
\end{equation}
for all relevant $2n$-quasiparticle states.  For example, in BCC with single and double excitations, we suppose that $\mathcal{T}(\varphi)$ satisfies the angle-dependent BCC equations for all quasi-particle excitations $\langle \Phi^{\mu} |$ corresponding to $n=1$ and for $n=2$.  Using the relation expressed in Eq. \ref{Eq:Hz} we see that equivalently
\begin{equation}
\langle \Phi^{\mu} | e^{-\mathcal{T}(\varphi)} \, e^{-Y(\varphi)} \, e^{-X(\varphi)} \, H \, e^{X(\varphi)} \, e^{Y(\varphi)}  \, e^{\mathcal{T}(\varphi)} |\Phi \rangle = 0.
\label{Eqn:TMHBCC1}
\end{equation}
Because $Y(\varphi)$ preserves the number of quasiparticles, the set of states $\langle \Phi^{\mu} |$ and $\langle \Phi^{\mu} | \exp(-Y(\varphi))$ span the same space, so satisfaction of Eq. \ref{Eqn:TMHBCC1} means that
\begin{equation}
\langle \Phi^{\mu} | e^{-\hat{\mathcal{T}}(\varphi)} \, H \, e^{\hat{\mathcal{T}}(\varphi)} |\Phi \rangle = 0
\end{equation}
is also satisfied.  In other words, if $\mathcal{T}(\varphi)$ satisfies the angle-dependent BCC equations for a given excitation level, then $\hat{\mathcal{T}}(\varphi)$ satisfies the unrotated BCC equations for the same excitation level, i.e.
\begin{equation}
\hat{\mathcal{T}}(\varphi) = \mathcal{T}(0).
\end{equation}
From Eq. \ref{Eqn:LittleH} the reduced Hamiltonian kernel $h(\varphi)$ is then
\begin{equation}
h(\varphi) = \langle \Phi| H \, e^{\mathcal{T}(0)} |\Phi\rangle = h(0)
\end{equation}
and the projected energy of  Eq. \ref{projeigenequatkernelsB1} is simply
\begin{equation}
E^A = h(0) \, \frac{\int_{0}^{2\pi} \!d\varphi \, e^{-i\text{A}\varphi}  \, {\cal N}(\varphi)}{\int_{0}^{2\pi} \!d\varphi \, e^{-i\text{A}\varphi}  \, {\cal N}(\varphi)} = h(0)
\end{equation}
which is nothing but the BCC energy associated with $\mathcal{T}(0)$.

To restate our result, regardless of the truncation of $U = \mathcal{T}(0)$, so long as is contains single excitations (because $\hat{\mathcal{T}}(\varphi)$ always contains single excitations) and so long as all BCC equations for the relevant excitation levels are satisfied, one solution of the angle-dependent BCC equations is such that the reduced Hamiltonian kernel is constant and equal to its value at $\varphi = 0$, which is the unprojected BCC energy.

\section{Pairing hamiltonian}
\label{PH}

The present section provides explicit expressions for the ingredients necessary to implement PBCS-CCSD theory to the pairing Hamiltonian problem. 

\subsection{BCS transformation}
\label{SecBCStransfo}

For sufficiently strong $G$, the variational solution of the pairing Hamiltonian within the manifold of product states develops a particle-number breaking solution $| \Phi \rangle$ with quasiparticle operators defined through the BCS transformation 
\begin{subequations}
\label{BCStransfo}
\begin{align}
    \beta^{\dagger}_p &= u_p c^{\dagger}_p - v_p c_{\bar{p}} \, ,
    \\
    \beta_{\bar{p}} &= u_p c_{\bar{p}} + v_p c^{\dagger}_{p} \, .
\end{align}
\end{subequations}
The BCS transformation corresponds to a simplified version of the generic Bogoliubov transformation introduced in Sec.~\ref{PCCansatz} characterized by $2\times 2$ block matrices of the form
\begin{subequations}
\label{BCSmatrices}
\begin{alignat}{2}
 V_{pp'}   &= V_{p \bar{p}} \, \delta_{p' \bar{p}} &&\equiv +v_{p} \, \delta_{\bar{p}p'}  \, , \\
 V_{\bar{p}p} &= -V_{p \bar{p}} &&= - v_{p}\, , \\
 U_{pp'}   &= U_{pp} \, \delta_{p' p} &&\equiv +u_{p} \, \delta_{pp'}  \, , \\
 U_{\bar{p}\bar{p}} &= +U_{pp} &&= + u_{p}    \, . 
\end{alignat}
\end{subequations}

\subsection{Normal-ordered Hamiltonian}
\label{QPME}

While the Hamiltonian is initially expressed in the single-particle basis (Eq.~\ref{Eqn:PairHam}), it is of interest to express it in the quasi-particle basis of the vacuum $| \Phi \rangle$. Inverting the BCS transformation~\ref{BCStransfo}
\begin{subequations}
\begin{align}
    c^{\dagger}_p &= u_p \beta^{\dagger}_p + v_p \beta_{\bar{p}} \, , \\
    c_{\bar{p}} &= u_p \beta_{\bar{p}} - v_p \beta^{\dagger}_{p} \, ,
\end{align}
\end{subequations}
one first expresses the pair operators in terms of quasi-particle ones defined through
\begin{subequations}
\begin{align}
\N_p &= \beta_{p}^\dagger \, \beta_{p}^{} + \beta_{\bar{p}}^\dagger \, \beta_{\bar{p}}^{} \, ,
\\
\P_p^\dagger &= \beta_{p}^\dagger \, \beta_{\bar{p}}^\dagger \, ,
\end{align}
\end{subequations}
according to
\begin{subequations}
\begin{align}
    N_p &= 2v_p^2 + 2u_pv_p\mathcal{P}^{\dagger}_p + (u_p^2-v_p^2)\mathcal{N}_p + 2u_pv_p\mathcal{P}_p \,  ,
    \\
    P^{\dagger}_p &= u_pv_p + u_p^2\mathcal{P}_p^{\dagger} - u_pv_p\mathcal{N}_p - v_p^2\mathcal{P}_p \, ,
    \\
    P_p &= u_pv_p - v_p^2\mathcal{P}_p^{\dagger} - u_pv_p\mathcal{N}_p + u_p^2\mathcal{P}_p \, .
\end{align}
\end{subequations}
Inserting these relations in Eq.~\ref{Eqn:PairHam}, the normal-ordered Hamiltonian under the form
\begin{align}
\label{NO_H}
    H &= H^{00} + \sum_p (H_p^{11}\mathcal{N}_p 
    + H_p^{02}\mathcal{P}^{\dagger}_p 
    + H_p^{20}\mathcal{P}_p)  \nonumber
    \\
    &+ \sum_{pq}( H_{pq}^{22}\mathcal{N}_p\mathcal{N}_q 
    + \tilde{H}_{pq}^{22}\mathcal{P}_p^{\dagger}\mathcal{P}_q
    + H_{pq}^{13}\mathcal{P}^{\dagger}_p\mathcal{N}_q
    + H_{pq}^{31}\mathcal{N}_p\mathcal{P}_q ) \nonumber
    \\
    &+ \sum_{p\neq q}(H_{pq}^{04}\mathcal{P}^{\dagger}_p\mathcal{P}^{\dagger}_q
    + H_{pq}^{40}\mathcal{P}_p\mathcal{P}_q) \, ,
\end{align}
where the matrix elements associated with the various normal-ordered terms are given by
\begin{subequations}
\begin{align}
    H^{00} &= 2\sum_p(\epsilon_p-\lambda)v_p^2  \nonumber \\
    &- G\big(\sum_{p}u_pv_p\big)^2 - G\sum_pv_p^4 \, ,
    \\
    H^{11}_p &= (\epsilon_p-\lambda)(u_p^2-v_p^2) \nonumber \\
    &+ 2Gu_pv_p \sum_{q}u_qv_q + Gv_p^4  \, ,
    \\
    H^{20}_p &= H^{02}_p = 2(\epsilon_p-\lambda) u_pv_p \nonumber \\
    & - G(u_p^2-v_p^2)\sum_{q}u_qv_q
    - 2Gu_pv_p^3 \, ,
    \\
    H^{22}_{pq} &= -Gu_pv_pu_qv_q \, ,
    \\
    \tilde{H}^{22}_{pq} &= -G(u_p^2u_q^2 + v_p^2v_q^2) \, ,
    \\
    H^{31}_{qp} &= H^{13}_{pq} = G(u_p^2-v_p^2)u_qv_q \, ,
    \\
    H^{40}_{pq} &= H^{04}_{pq} = G\frac{u_p^2v_q^2 + v_p^2u_q^2}{2} \, .
\end{align}
\end{subequations}

\subsection{Thouless transformation}

The gauge-rotated BCS state $\langle \Phi(\varphi) |$ is expressed via the Thouless transformation of the unrotated one (Eq.~\ref{thoulessbetweenbothvacua}). Given the BCS transformation (Eq.~\ref{BCSmatrices}), the plain overlap between both states reads as
\begin{align}
\langle \Phi(\varphi) | \Phi \rangle & = \prod_p (u_p^2 + e^{2i\varphi}v^2_p)\, , \label{thoulesstransfoBCS2} 
\end{align}
whereas the transition BCS transformation (Eq.~\ref{transfobogotransition}) reads as
\begin{subequations}
\label{transfobogotransitionBCS}
\begin{align}
M_{pp'}(\varphi) & \equiv e^{-i\varphi} (u_p^2 + e^{2i\varphi} v_p^2)\delta_{pp'} \, , \label{transfobogotransitionBCSM}  \\
N_{pp'}(\varphi) & \equiv e^{-i\varphi} u_p v_p  (e^{2i\varphi}-1) \delta_{\bar{p}p'}  \, . \label{transfobogotransitionBCSN} 
\end{align}
\end{subequations}
The latter matrices allows one to write the Thouless matrix (Eq.~\ref{thoulessoprot}) under the form
\begin{align}
Z^{02}_{pp'}(\varphi) & = \frac{u_{p} v_{p}\big( e^{2i\varphi} - 1 \big)}{u_{p}^2 + e^{2i\varphi}v_{p}^2 }  \delta_{\bar{p}p'} 
\equiv z_{p}(\varphi) \delta_{\bar{p}p'} \, , \label{thoulesstransfoBCS1}  
\end{align}
such that the Thouless operator reduces to
\begin{equation}
\label{thoulessoprotBCS}
Z(\varphi) \equiv \sum_{p} z_{p}(\varphi)  \P_p \, .
\end{equation}

\subsection{Similarity transformation}

The similarity transformation by the Thouless operator was detailed in App.~\ref{Sectransforop}. In the context of the pairing hamiltonian problem, one first needs to transform the pair quasiparticle operators according to
\begin{align}
    e^{Z(\varphi)} \mathcal{P}_p e^{-Z(\varphi)} &= \mathcal{P}_p \, , \nonumber
    \\
    e^{Z(\varphi)} \mathcal{N}_p e^{-Z(\varphi)} &= \mathcal{N}_p + 2z_{p}(\varphi)\mathcal{P}_p \, ,
    \\
    e^{Z(\varphi)} \mathcal{P}_p^{\dagger} e^{-Z(\varphi)} &= \mathcal{P}_p^{\dagger}
    + z_{p}(\varphi)(1-\mathcal{N}_p) - z^2_{p}(\varphi)\mathcal{P}_p \, . \nonumber
\end{align}
With this result at hand, one obtains the similarity-transformed Hamiltonian under the form
\begin{align}
\label{NO_H_transfo}
    H_{Z}(\varphi) &\equiv H^{00}(\varphi) \nonumber \\
    &+ \sum_p (H_p^{11}(\varphi)\mathcal{N}_p 
    + H_p^{02}(\varphi)\mathcal{P}^{\dagger}_p 
    + H_p^{20}(\varphi)\mathcal{P}_p)  \nonumber
    \\
    &+ \sum_{pq} \big( H_{pq}^{22}(\varphi)\mathcal{N}_p\mathcal{N}_q 
    + \tilde{H}_{pq}^{22}(\varphi)\mathcal{P}_p^{\dagger}\mathcal{P}_q  \nonumber
    \\
    &
    \hspace{0.7cm}+ H_{pq}^{13}(\varphi)\mathcal{P}^{\dagger}_p\mathcal{N}_q
    + H_{pq}^{31}(\varphi)\mathcal{N}_p\mathcal{P}_q \big) \nonumber
    \\
    &+ \sum_{p\neq q}(H_{pq}^{04}(\varphi)\mathcal{P}^{\dagger}_p\mathcal{P}^{\dagger}_q
    + H_{pq}^{40}(\varphi)\mathcal{P}_p\mathcal{P}_q) \, ,
\end{align}
where the matrix elements are given by
\begin{subequations}
\begin{align}
    H^{00}(\varphi) &= H^{00} + \sum_p H^{02}_p z_{p}(\varphi) \nonumber \\
    &\,\,\,+ \sum_{pq} H^{04}_{pq} z_{p}(\varphi)z_{q}(\varphi)
    - \sum_{p} H^{04}_{pp}z^2_{p}(\varphi)    \, ,
    \\
    H_p^{11}(\varphi) &= H^{11}_p - H^{02}_{p} z_{p}(\varphi) \nonumber \\
    &\,\,\,- \sum_{q}(H^{04}_{pq}+H^{04}_{qp}) z_{p}(\varphi) z_{q}(\varphi) \nonumber \\
    &\,\,\,+ H^{04}_{pp} z^2_{p}(\varphi) + \sum_{q} H^{13}_{qp} z_{q}(\varphi)  \, ,
    \\
    H_p^{02}(\varphi) &= H^{02}_{p} - 2H^{04}_{pp} z_{p}(\varphi) \nonumber \\
    &\,\,\,+ \sum_{q}(H^{04}_{pq}+H^{04}_{qp}) z_{q}(\varphi)   \, ,
    \\
     H_p^{20}(\varphi) &= H^{20}_{p} + 2H^{11}_{p} z_{p}(\varphi) - H^{02}_{p} z^2_{p}(\varphi)
        \nonumber \\
    &\,\,\,+ 4H^{22}_{pp} z_{p}(\varphi) + \sum_{q}\tilde{H}^{22}_{qp} z_{q}(\varphi) \nonumber \\
    &\,\,\, - \sum_{q}(H^{04}_{pq}+H^{04}_{qp}) z^2_{p}(\varphi) z_{q}(\varphi) + 2H^{04}_{pp} z^3_{p}(\varphi)\nonumber \\
    &\,\,\,+ \sum_{q}2H^{13}_{qp} z_{p}(\varphi) z_{q}(\varphi) - 2H^{13}_{pp} z^2_{p}(\varphi)  \, ,
    \\
    H_{pq}^{22}(\varphi) &= H^{22}_{pq} + H^{04}_{pq} z_{p}(\varphi) z_{q}(\varphi) - H^{13}_{pq} z_{p}(\varphi)  \, ,
    \\
    \tilde{H}_{pq}^{22}(\varphi) &= \tilde{H}^{22}_{pq} - (H^{04}_{pq}+H^{04}_{qp}) z^2_{p}(\varphi) \nonumber \\
    &\,\,\,+ 2H^{13}_{pq} z_{q}(\varphi)    \, ,
    \\
    H_{pq}^{04}(\varphi) &= H_{pq}^{04}  \, ,
    \\
    H_{pq}^{40}(\varphi) &= H_{pq}^{40} + 4H^{22}_{pq} z_{p}(\varphi) z_{q}(\varphi) \nonumber \\
    &\,\,\,- \tilde{H}^{22}_{pq} z^2_{p}(\varphi)   + H^{04}_{pq} z^2_{p}(\varphi) z^2_{q}(\varphi) \nonumber \\
    &\,\,\,- 2H^{13}_{pq} z^2_{p}(\varphi) z_{q}(\varphi) + 2H^{31}_{pq} z_{p}(\varphi)    \, , 
    \\
    H_{pq}^{13}(\varphi) &= H^{13}_{pq} - (H^{04}_{pq}+H^{04}_{qp}) z_{q}(\varphi)  \, ,
    \\
    H_{pq}^{31}(\varphi) &= H^{31}_{pq} + 2(H^{22}_{pq}+H^{22}_{qp}) z_{q}(\varphi) \nonumber \\
    &\,\,\,- \tilde{H}^{22}_{pq} z_{p}(\varphi) + 
    (H^{04}_{pq}+H^{04}_{qp}) z_{p}(\varphi) z^2_{q}(\varphi) \nonumber \\
    &\,\,\,- 2H^{13}_{pq} z_{p}(\varphi) z_{q}(\varphi) - H^{13}_{qp} z^2_{q}(\varphi) \, .
\end{align}
\end{subequations}

\subsection{Hamiltonian and norm kernels}

Unexcited and excited Hamiltonian and norm kernels introduced in Eqs.~\ref{kernelsA} and~\ref{excitkernelsB} are the main building blocks to solve for the amplitude equations and compute the projected energy. Writing them in terms of their connected counterparts, they read as 
\begin{subequations}
\begin{align}
    \tilde{\mathcal{N}}_p(\varphi) &= n_p(\varphi) \, \mathcal{N}(\varphi)  \, ,
    \\
    \tilde{\mathcal{N}}_{pq}(\varphi) &= n_{pq}(\varphi) \, \mathcal{N}(\varphi)  \, ,
    \\
    \tilde{\mathcal{H}}_p(\varphi) &= \Big( h_p(\varphi) 
    + h(\varphi)n_p(\varphi)\Big) \mathcal{N}(\varphi)  \, ,
    \\
    \tilde{\mathcal{H}}_{pq}(\varphi) &= \Big( h_{pq}(\varphi) 
    + h_p(\varphi) n_q(\varphi) \nonumber \\
    &\,\,\,\,\,\,+ h_q(\varphi) n_p(\varphi)
    + h(\varphi)n_{pq}(\varphi)\Big) \mathcal{N}(\varphi)  \, ,
\end{align}
\end{subequations}
with
\begin{subequations}
\begin{align}
    \mathcal{N}(\varphi) &= \langle \Phi(\varphi) | \Phi \rangle \, e^{W_0(\varphi)} \, ,
    \\
    n_p(\varphi) &= \langle \Phi | \mathcal{P}_p e^{\mathcal{T}(\varphi)} | \Phi \rangle \nonumber \\
    &
    = W_p(\varphi) \, ,
    \\
    n_{pq}(\varphi) &= \langle \Phi | \mathcal{P}_q \mathcal{P}_p e^{\mathcal{T}(\varphi)} | \Phi \rangle\nonumber \\
    &
    = W_{pq}(\varphi) + W_p(\varphi)W_q(\varphi) \, ,
\end{align}
\end{subequations}
and
\begin{subequations}
\begin{align}
    h(\varphi) &= \langle \Phi | H_Z(\varphi) e^{\mathcal{T}(\varphi)} | \Phi \rangle_C \nonumber \\
    &
    = \langle \Phi | \bar{H}_Z(\varphi) | \Phi \rangle \, ,
    \\
    h_p(\varphi) &= \langle \Phi | \mathcal{P}_p H_Z(\varphi) e^{\mathcal{T}(\varphi)} | \Phi \rangle_C \nonumber \\
    &
    = \langle \Phi | \mathcal{P}_p \bar{H}_Z(\varphi) | \Phi \rangle \, ,
    \\
    h_{pq}(\varphi) &= \langle \Phi | \mathcal{P}_q \mathcal{P}_p H_Z(\varphi) e^{\mathcal{T}(\varphi)} | \Phi \rangle_C \nonumber \\
    &
    = \langle \Phi | \mathcal{P}_q \bar{H}_Z(\varphi) | \Phi \rangle \, ,
\end{align}
\end{subequations}
where $W_p(\varphi)$ and $W_{pq}(\varphi)$ denote single and double gauge-rotated cluster amplitudes, respectively. The connected kernels of the Hamiltonian $h(\varphi)$, $h_p(\varphi)$ and $h_{pq}(\varphi)$ are nothing but energy, single and double BCC residuals for the doubly similarity-transformed Hamiltonian $\bar{H}_Z(\varphi)$ such that their expressions are easily obtained from BCC ones~\cite{Henderson:2014vka} at the price of replacing the initial matrix elements of $H$ by those of $H_Z(\varphi)$.

\subsection{Differential equations}

In Eq.~\ref{ntuply1storderdiffeq0}, the differential equations satisfied by gauge-rotated cluster amplitudes were provided. Specifying them to the pairing Hamiltonian problem and going up to quadruple amplitudes, they read as 
\begin{subequations}
\begin{align}
    \frac{\mathrm{d}W_0(\varphi)}{\mathrm{d}\varphi} &= i\sum_p A^{02}_p(\varphi) W_p(\varphi) \, ,
    \\
    \frac{\mathrm{d}W_p(\varphi)}{\mathrm{d}\varphi} 
    &= -iA^{02}_p(\varphi)W_p^2(\varphi) \nonumber \\
    & \hspace{0.5cm}+ i\sum_p A^{02}_q(\varphi) W_{pq}(\varphi) \, ,
    \\
    \frac{\mathrm{d}W_{pq}(\varphi)}{\mathrm{d}\varphi} 
    & = -2 i\,\textbf{p}(pq) A^{02}_p(\varphi)W_{p}(\varphi)W_{pq}(\varphi)\nonumber \\
    & \hspace{0.5cm} + i\sum_r A^{02}_r(\varphi) W_{pqr}(\varphi) \, ,
    \\
    \frac{\mathrm{d}W_{pqr}(\varphi)}{\mathrm{d}\varphi} 
    & = -2 i\,\textbf{p}(pqr) A^{02}_p(\varphi)W_{pq}(\varphi)W_{pr}(\varphi) \nonumber \\
    & \hspace{0.5cm}-2 i\,\textbf{p}(pqr) A^{02}_p(\varphi)W_p(\varphi) W_{pqr}(\varphi)\nonumber \\
    & \hspace{0.5cm}
    + i\sum_s A^{02}_s(\varphi) W_{pqrs}(\varphi) \, ,
    \\
    \frac{\mathrm{d}W_{pqrs}(\varphi)}{\mathrm{d}\varphi} 
    & = -2 i\,\textbf{p}(pqrs) A^{02}_p(\varphi)W_{pq}(\varphi)W_{prs}(\varphi) \nonumber \\
    & \hspace{0.5cm}-2 i \,\textbf{p}(pqrs) A^{02}_p(\varphi)W_p(\varphi) W_{pqrs}(\varphi)\nonumber \\
    & \hspace{0.5cm}
    + i\sum_t A^{02}_t(\varphi) W_{pqrst}(\varphi) \, ,
\end{align}
\end{subequations}
where $\textbf{p}(pq\cdots)$ permutes the indices to create terms that are distinct when taking 
the symmetries of the indices of $W(\varphi)$ into account. From the explicit expression, it is clear that
$dW_k(\varphi)$ contains contributions from $W_{k+1}(\varphi)$. By truncating at a certain excitation level, these
equations can be decoupled, and the low order amplitudes can be obtained from an ODE solver.

\bibliography{P-BCS-CC}

\begin{thebibliography}{30}%
\makeatletter
\providecommand \@ifxundefined [1]{%
 \@ifx{#1\undefined}
}%
\providecommand \@ifnum [1]{%
 \ifnum #1\expandafter \@firstoftwo
 \else \expandafter \@secondoftwo
 \fi
}%
\providecommand \@ifx [1]{%
 \ifx #1\expandafter \@firstoftwo
 \else \expandafter \@secondoftwo
 \fi
}%
\providecommand \natexlab [1]{#1}%
\providecommand \enquote  [1]{``#1''}%
\providecommand \bibnamefont  [1]{#1}%
\providecommand \bibfnamefont [1]{#1}%
\providecommand \citenamefont [1]{#1}%
\providecommand \href@noop [0]{\@secondoftwo}%
\providecommand \href [0]{\begingroup \@sanitize@url \@href}%
\providecommand \@href[1]{\@@startlink{#1}\@@href}%
\providecommand \@@href[1]{\endgroup#1\@@endlink}%
\providecommand \@sanitize@url [0]{\catcode `\\12\catcode `\$12\catcode
  `\&12\catcode `\#12\catcode `\^12\catcode `\_12\catcode `\%12\relax}%
\providecommand \@@startlink[1]{}%
\providecommand \@@endlink[0]{}%
\providecommand \url  [0]{\begingroup\@sanitize@url \@url }%
\providecommand \@url [1]{\endgroup\@href {#1}{\urlprefix }}%
\providecommand \urlprefix  [0]{URL }%
\providecommand \Eprint [0]{\href }%
\providecommand \doibase [0]{http://dx.doi.org/}%
\providecommand \selectlanguage [0]{\@gobble}%
\providecommand \bibinfo  [0]{\@secondoftwo}%
\providecommand \bibfield  [0]{\@secondoftwo}%
\providecommand \translation [1]{[#1]}%
\providecommand \BibitemOpen [0]{}%
\providecommand \bibitemStop [0]{}%
\providecommand \bibitemNoStop [0]{.\EOS\space}%
\providecommand \EOS [0]{\spacefactor3000\relax}%
\providecommand \BibitemShut  [1]{\csname bibitem#1\endcsname}%
\let\auto@bib@innerbib\@empty
\bibitem [{\citenamefont {Coester}(1958)}]{Coester1958}%
  \BibitemOpen
  \bibfield  {author} {\bibinfo {author} {\bibfnamefont {F.}~\bibnamefont
  {Coester}},\ }\href@noop {} {\bibfield  {journal} {\bibinfo  {journal}
  {Nuclear Phys.}\ }\textbf {\bibinfo {volume} {7}},\ \bibinfo {pages} {421}
  (\bibinfo {year} {1958})}\BibitemShut {NoStop}%
\bibitem [{\citenamefont {\v{C}\'i\v{z}ek}(1966)}]{Cizek1966}%
  \BibitemOpen
  \bibfield  {author} {\bibinfo {author} {\bibfnamefont {J.}~\bibnamefont
  {\v{C}\'i\v{z}ek}},\ }\href@noop {} {\bibfield  {journal} {\bibinfo
  {journal} {J. Chem. Phys.}\ }\textbf {\bibinfo {volume} {45}},\ \bibinfo
  {pages} {4256} (\bibinfo {year} {1966})}\BibitemShut {NoStop}%
\bibitem [{\citenamefont {Paldus}\ and\ \citenamefont {Li}(2007)}]{Paldus1999}%
  \BibitemOpen
  \bibfield  {author} {\bibinfo {author} {\bibfnamefont {J.}~\bibnamefont
  {Paldus}}\ and\ \bibinfo {author} {\bibfnamefont {X.~Z.}\ \bibnamefont
  {Li}},\ }\href@noop {} {\bibfield  {journal} {\bibinfo  {journal} {Adv. Chem.
  Phys.}\ }\textbf {\bibinfo {volume} {110}},\ \bibinfo {pages} {1} (\bibinfo
  {year} {2007})}\BibitemShut {NoStop}%
\bibitem [{\citenamefont {Bartlett}\ and\ \citenamefont
  {Musia{\l}}(2007)}]{Bartlett2007}%
  \BibitemOpen
  \bibfield  {author} {\bibinfo {author} {\bibfnamefont {R.~J.}\ \bibnamefont
  {Bartlett}}\ and\ \bibinfo {author} {\bibfnamefont {M.}~\bibnamefont
  {Musia{\l}}},\ }\href@noop {} {\bibfield  {journal} {\bibinfo  {journal}
  {Rev. Mod. Phys.}\ }\textbf {\bibinfo {volume} {79}},\ \bibinfo {pages} {291}
  (\bibinfo {year} {2007})}\BibitemShut {NoStop}%
\bibitem [{\citenamefont {Shavitt}\ and\ \citenamefont
  {Bartlett}(2009)}]{ShavittBartlett}%
  \BibitemOpen
  \bibfield  {author} {\bibinfo {author} {\bibfnamefont {I.}~\bibnamefont
  {Shavitt}}\ and\ \bibinfo {author} {\bibfnamefont {R.~J.}\ \bibnamefont
  {Bartlett}},\ }\href@noop {} {\emph {\bibinfo {title} {Many-Body Methods in
  Chemistry and Physics}}}\ (\bibinfo  {publisher} {Cambridge University
  Press},\ \bibinfo {address} {New York},\ \bibinfo {year} {2009})\ \bibinfo
  {note} {and references contained therein}\BibitemShut {NoStop}%
\bibitem [{\citenamefont {Ui}\ and\ \citenamefont {Takeda}(1983)}]{ui83a}%
  \BibitemOpen
  \bibfield  {author} {\bibinfo {author} {\bibfnamefont {H.}~\bibnamefont
  {Ui}}\ and\ \bibinfo {author} {\bibfnamefont {G.}~\bibnamefont {Takeda}},\
  }\href@noop {} {\bibfield  {journal} {\bibinfo  {journal} {Prog. Theor.
  Phys.}\ }\textbf {\bibinfo {volume} {70}},\ \bibinfo {pages} {176} (\bibinfo
  {year} {1983})}\BibitemShut {NoStop}%
\bibitem [{\citenamefont {Yannouleas}\ and\ \citenamefont
  {Landman}(2007)}]{yannouleas07a}%
  \BibitemOpen
  \bibfield  {author} {\bibinfo {author} {\bibfnamefont {C.}~\bibnamefont
  {Yannouleas}}\ and\ \bibinfo {author} {\bibfnamefont {U.}~\bibnamefont
  {Landman}},\ }\href@noop {} {\bibfield  {journal} {\bibinfo  {journal} {Rep.
  Prog. Phys.}\ }\textbf {\bibinfo {volume} {70}},\ \bibinfo {pages} {2067}
  (\bibinfo {year} {2007})}\BibitemShut {NoStop}%
\bibitem [{\citenamefont {Papenbrock}\ and\ \citenamefont
  {Weidenmueller}(2014)}]{Papenbrock:2013cra}%
  \BibitemOpen
  \bibfield  {author} {\bibinfo {author} {\bibfnamefont {T.}~\bibnamefont
  {Papenbrock}}\ and\ \bibinfo {author} {\bibfnamefont {H.}~\bibnamefont
  {Weidenmueller}},\ }\href@noop {} {\bibfield  {journal} {\bibinfo  {journal}
  {Phys. Rev.}\ }\textbf {\bibinfo {volume} {C89}},\ \bibinfo {pages} {014334}
  (\bibinfo {year} {2014})}\BibitemShut {NoStop}%
\bibitem [{\citenamefont {L\"owdin}(1955)}]{Lowdin55c}%
  \BibitemOpen
  \bibfield  {author} {\bibinfo {author} {\bibfnamefont {P.-O.}\ \bibnamefont
  {L\"owdin}},\ }\href@noop {} {\bibfield  {journal} {\bibinfo  {journal}
  {Phys. Rev.}\ }\textbf {\bibinfo {volume} {97}},\ \bibinfo {pages} {1509}
  (\bibinfo {year} {1955})}\BibitemShut {NoStop}%
\bibitem [{\citenamefont {Pauncz}(1967)}]{pauncz1967}%
  \BibitemOpen
  \bibfield  {author} {\bibinfo {author} {\bibfnamefont {R.}~\bibnamefont
  {Pauncz}},\ }\href@noop {} {\emph {\bibinfo {title} {Alternant Molecular
  Orbital Method}}}\ (\bibinfo  {publisher} {W.B. Saunders Company},\ \bibinfo
  {address} {Philadelphia},\ \bibinfo {year} {1967})\BibitemShut {NoStop}%
\bibitem [{\citenamefont {Mayer}(1980)}]{mayer1980}%
  \BibitemOpen
  \bibfield  {author} {\bibinfo {author} {\bibfnamefont {I.}~\bibnamefont
  {Mayer}},\ }\href@noop {} {\bibfield  {journal} {\bibinfo  {journal} {Adv.
  Quantum Chem.}\ }\textbf {\bibinfo {volume} {12}},\ \bibinfo {pages} {189}
  (\bibinfo {year} {1980})}\BibitemShut {NoStop}%
\bibitem [{\citenamefont {Ring}\ and\ \citenamefont {Schuck}(1980)}]{ring80a}%
  \BibitemOpen
  \bibfield  {author} {\bibinfo {author} {\bibfnamefont {P.}~\bibnamefont
  {Ring}}\ and\ \bibinfo {author} {\bibfnamefont {P.}~\bibnamefont {Schuck}},\
  }\href@noop {} {\emph {\bibinfo {title} {The Nuclear Many-Body Problem}}}\
  (\bibinfo  {publisher} {Springer-Verlag},\ \bibinfo {address} {New-York},\
  \bibinfo {year} {1980})\BibitemShut {NoStop}%
\bibitem [{\citenamefont {Blaizot}\ and\ \citenamefont
  {Ripka}(1985)}]{Blaizot85}%
  \BibitemOpen
  \bibfield  {author} {\bibinfo {author} {\bibfnamefont {J.-P.}\ \bibnamefont
  {Blaizot}}\ and\ \bibinfo {author} {\bibfnamefont {G.}~\bibnamefont
  {Ripka}},\ }\href@noop {} {\emph {\bibinfo {title} {Quantum Theory of Finite
  Systems}}}\ (\bibinfo  {publisher} {The MIT Press},\ \bibinfo {address}
  {Cambridge, MA},\ \bibinfo {year} {1985})\BibitemShut {NoStop}%
\bibitem [{\citenamefont {Bender}\ \emph {et~al.}(2003)\citenamefont {Bender},
  \citenamefont {Heenen},\ and\ \citenamefont {Reinhard}}]{bender03b}%
  \BibitemOpen
  \bibfield  {author} {\bibinfo {author} {\bibfnamefont {M.}~\bibnamefont
  {Bender}}, \bibinfo {author} {\bibfnamefont {P.-H.}\ \bibnamefont {Heenen}},
  \ and\ \bibinfo {author} {\bibfnamefont {P.-G.}\ \bibnamefont {Reinhard}},\
  }\href@noop {} {\bibfield  {journal} {\bibinfo  {journal} {Rev. Mod. Phys.}\
  }\textbf {\bibinfo {volume} {75}},\ \bibinfo {pages} {121} (\bibinfo {year}
  {2003})}\BibitemShut {NoStop}%
\bibitem [{\citenamefont {Schmid}(2004)}]{Schmid2004}%
  \BibitemOpen
  \bibfield  {author} {\bibinfo {author} {\bibfnamefont {K.~W.}\ \bibnamefont
  {Schmid}},\ }\href@noop {} {\bibfield  {journal} {\bibinfo  {journal} {Prog.
  Part. Nucl. Phys.}\ }\textbf {\bibinfo {volume} {52}},\ \bibinfo {pages}
  {565} (\bibinfo {year} {2004})}\BibitemShut {NoStop}%
\bibitem [{\citenamefont {Jim\'enez-Hoyos}\ \emph {et~al.}(2012)\citenamefont
  {Jim\'enez-Hoyos}, \citenamefont {Henderson}, \citenamefont {Tsuchimochi},\
  and\ \citenamefont {Scuseria}}]{PHF}%
  \BibitemOpen
  \bibfield  {author} {\bibinfo {author} {\bibfnamefont {C.~A.}\ \bibnamefont
  {Jim\'enez-Hoyos}}, \bibinfo {author} {\bibfnamefont {T.~M.}\ \bibnamefont
  {Henderson}}, \bibinfo {author} {\bibfnamefont {T.}~\bibnamefont
  {Tsuchimochi}}, \ and\ \bibinfo {author} {\bibfnamefont {G.~E.}\ \bibnamefont
  {Scuseria}},\ }\href@noop {} {\bibfield  {journal} {\bibinfo  {journal} {J.
  Chem. Phys.}\ }\textbf {\bibinfo {volume} {136}},\ \bibinfo {pages} {164109}
  (\bibinfo {year} {2012})}\BibitemShut {NoStop}%
\bibitem [{\citenamefont {Piecuch}\ \emph {et~al.}(1996)\citenamefont
  {Piecuch}, \citenamefont {Tobo{\l}a},\ and\ \citenamefont
  {Paldus}}]{Piecuch1996}%
  \BibitemOpen
  \bibfield  {author} {\bibinfo {author} {\bibfnamefont {P.}~\bibnamefont
  {Piecuch}}, \bibinfo {author} {\bibfnamefont {R.}~\bibnamefont {Tobo{\l}a}},
  \ and\ \bibinfo {author} {\bibfnamefont {J.}~\bibnamefont {Paldus}},\
  }\href@noop {} {\bibfield  {journal} {\bibinfo  {journal} {Phys. Rev. A}\
  }\textbf {\bibinfo {volume} {54}},\ \bibinfo {pages} {1210} (\bibinfo {year}
  {1996})}\BibitemShut {NoStop}%
\bibitem [{\citenamefont {Qiu}\ \emph {et~al.}(2017{\natexlab{a}})\citenamefont
  {Qiu}, \citenamefont {Henderson},\ and\ \citenamefont {Scuseria}}]{Qiu2017}%
  \BibitemOpen
  \bibfield  {author} {\bibinfo {author} {\bibfnamefont {Y.}~\bibnamefont
  {Qiu}}, \bibinfo {author} {\bibfnamefont {T.~M.}\ \bibnamefont {Henderson}},
  \ and\ \bibinfo {author} {\bibfnamefont {G.~E.}\ \bibnamefont {Scuseria}},\
  }\href@noop {} {\bibfield  {journal} {\bibinfo  {journal} {J. Chem. Phys.}\
  }\textbf {\bibinfo {volume} {146}},\ \bibinfo {pages} {184105} (\bibinfo
  {year} {2017}{\natexlab{a}})}\BibitemShut {NoStop}%
\bibitem [{\citenamefont {Duguet}(2015)}]{Duguet:2014jja}%
  \BibitemOpen
  \bibfield  {author} {\bibinfo {author} {\bibfnamefont {T.}~\bibnamefont
  {Duguet}},\ }\href@noop {} {\bibfield  {journal} {\bibinfo  {journal} {J.
  Phys.}\ }\textbf {\bibinfo {volume} {G42}},\ \bibinfo {pages} {025107}
  (\bibinfo {year} {2015})}\BibitemShut {NoStop}%
\bibitem [{\citenamefont {Duguet}\ and\ \citenamefont
  {Signoracci}(2017)}]{Duguet:2015yle}%
  \BibitemOpen
  \bibfield  {author} {\bibinfo {author} {\bibfnamefont {T.}~\bibnamefont
  {Duguet}}\ and\ \bibinfo {author} {\bibfnamefont {A.}~\bibnamefont
  {Signoracci}},\ }\href@noop {} {\bibfield  {journal} {\bibinfo  {journal} {J.
  Phys.}\ }\textbf {\bibinfo {volume} {G44}},\ \bibinfo {pages} {015103}
  (\bibinfo {year} {2017})}\BibitemShut {NoStop}%
\bibitem [{\citenamefont {Qiu}\ \emph {et~al.}(2017{\natexlab{b}})\citenamefont
  {Qiu}, \citenamefont {Henderson}, \citenamefont {Zhao},\ and\ \citenamefont
  {Scuseria}}]{qiu17a}%
  \BibitemOpen
  \bibfield  {author} {\bibinfo {author} {\bibfnamefont {Y.}~\bibnamefont
  {Qiu}}, \bibinfo {author} {\bibfnamefont {T.~M.}\ \bibnamefont {Henderson}},
  \bibinfo {author} {\bibfnamefont {J.}~\bibnamefont {Zhao}}, \ and\ \bibinfo
  {author} {\bibfnamefont {G.~E.}\ \bibnamefont {Scuseria}},\ }\href@noop {}
  {\bibfield  {journal} {\bibinfo  {journal} {J. Chem. Phys.}\ }\textbf
  {\bibinfo {volume} {147}},\ \bibinfo {pages} {064111} (\bibinfo {year}
  {2017}{\natexlab{b}})}\BibitemShut {NoStop}%
\bibitem [{\citenamefont {Tsuchimochi}\ and\ \citenamefont
  {Ten-no}(2017)}]{Tsuchimochi2017}%
  \BibitemOpen
  \bibfield  {author} {\bibinfo {author} {\bibfnamefont {T.}~\bibnamefont
  {Tsuchimochi}}\ and\ \bibinfo {author} {\bibfnamefont {S.}~\bibnamefont
  {Ten-no}},\ }\href {\doibase 10.1021/acs.jctc.7b00073} {\bibfield  {journal}
  {\bibinfo  {journal} {J. Chem. Theory Comput.}\ }\textbf {\bibinfo {volume}
  {13}},\ \bibinfo {pages} {1667} (\bibinfo {year} {2017})}\BibitemShut
  {NoStop}%
\bibitem [{\citenamefont {Henderson}\ \emph {et~al.}(2014)\citenamefont
  {Henderson}, \citenamefont {Dukelsky}, \citenamefont {Scuseria},
  \citenamefont {Signoracci},\ and\ \citenamefont
  {Duguet}}]{Henderson:2014vka}%
  \BibitemOpen
  \bibfield  {author} {\bibinfo {author} {\bibfnamefont {T.~M.}\ \bibnamefont
  {Henderson}}, \bibinfo {author} {\bibfnamefont {J.}~\bibnamefont {Dukelsky}},
  \bibinfo {author} {\bibfnamefont {G.~E.}\ \bibnamefont {Scuseria}}, \bibinfo
  {author} {\bibfnamefont {A.}~\bibnamefont {Signoracci}}, \ and\ \bibinfo
  {author} {\bibfnamefont {T.}~\bibnamefont {Duguet}},\ }\href@noop {}
  {\bibfield  {journal} {\bibinfo  {journal} {Phys. Rev.}\ }\textbf {\bibinfo
  {volume} {C89}},\ \bibinfo {pages} {054305} (\bibinfo {year}
  {2014})}\BibitemShut {NoStop}%
\bibitem [{\citenamefont {Signoracci}\ \emph {et~al.}(2015)\citenamefont
  {Signoracci}, \citenamefont {Duguet}, \citenamefont {Hagen},\ and\
  \citenamefont {Jansen}}]{Signoracci:2014dia}%
  \BibitemOpen
  \bibfield  {author} {\bibinfo {author} {\bibfnamefont {A.}~\bibnamefont
  {Signoracci}}, \bibinfo {author} {\bibfnamefont {T.}~\bibnamefont {Duguet}},
  \bibinfo {author} {\bibfnamefont {G.}~\bibnamefont {Hagen}}, \ and\ \bibinfo
  {author} {\bibfnamefont {G.}~\bibnamefont {Jansen}},\ }\href {\doibase
  10.1103/PhysRevC.91.064320} {\bibfield  {journal} {\bibinfo  {journal} {Phys.
  Rev.}\ }\textbf {\bibinfo {volume} {C91}},\ \bibinfo {pages} {064320}
  (\bibinfo {year} {2015})},\ \Eprint {http://arxiv.org/abs/1412.2696}
  {arXiv:1412.2696 [nucl-th]} \BibitemShut {NoStop}%
\bibitem [{\citenamefont {Qiu}\ \emph {et~al.}(2018)\citenamefont {Qiu},
  \citenamefont {Henderson}, \citenamefont {Zhao},\ and\ \citenamefont
  {Scuseria}}]{Qiu2018}%
  \BibitemOpen
  \bibfield  {author} {\bibinfo {author} {\bibfnamefont {Y.}~\bibnamefont
  {Qiu}}, \bibinfo {author} {\bibfnamefont {T.~M.}\ \bibnamefont {Henderson}},
  \bibinfo {author} {\bibfnamefont {J.}~\bibnamefont {Zhao}}, \ and\ \bibinfo
  {author} {\bibfnamefont {G.~E.}\ \bibnamefont {Scuseria}},\ }\href@noop {}
  {\bibfield  {journal} {\bibinfo  {journal} {J. Chem. Phys.}\ }\textbf
  {\bibinfo {volume} {149}},\ \bibinfo {pages} {164108} (\bibinfo {year}
  {2018})},\ \Eprint {http://arxiv.org/abs/1808.07972} {arXiv:1808.07972
  [chem-ph]} \BibitemShut {NoStop}%
\bibitem [{\citenamefont {Richardson}(1963)}]{Richardson1}%
  \BibitemOpen
  \bibfield  {author} {\bibinfo {author} {\bibfnamefont {R.~W.}\ \bibnamefont
  {Richardson}},\ }\href@noop {} {\bibfield  {journal} {\bibinfo  {journal}
  {Phys. Lett.}\ }\textbf {\bibinfo {volume} {3}},\ \bibinfo {pages} {277}
  (\bibinfo {year} {1963})}\BibitemShut {NoStop}%
\bibitem [{\citenamefont {Richardson}(1966)}]{Richardson4}%
  \BibitemOpen
  \bibfield  {author} {\bibinfo {author} {\bibfnamefont {R.~W.}\ \bibnamefont
  {Richardson}},\ }\href@noop {} {\bibfield  {journal} {\bibinfo  {journal}
  {Phys. Rev.}\ }\textbf {\bibinfo {volume} {141}},\ \bibinfo {pages} {949}
  (\bibinfo {year} {1966})}\BibitemShut {NoStop}%
\bibitem [{\citenamefont {Degroote}\ \emph {et~al.}(2016)\citenamefont
  {Degroote}, \citenamefont {Henderson}, \citenamefont {Zhao}, \citenamefont
  {Dukelsky},\ and\ \citenamefont {Scuseria}}]{Deg16}%
  \BibitemOpen
  \bibfield  {author} {\bibinfo {author} {\bibfnamefont {M.}~\bibnamefont
  {Degroote}}, \bibinfo {author} {\bibfnamefont {T.~M.}\ \bibnamefont
  {Henderson}}, \bibinfo {author} {\bibfnamefont {J.}~\bibnamefont {Zhao}},
  \bibinfo {author} {\bibfnamefont {J.}~\bibnamefont {Dukelsky}}, \ and\
  \bibinfo {author} {\bibfnamefont {G.~E.}\ \bibnamefont {Scuseria}},\
  }\href@noop {} {\bibfield  {journal} {\bibinfo  {journal} {Phys. Rev.}\
  }\textbf {\bibinfo {volume} {B93}},\ \bibinfo {pages} {125124} (\bibinfo
  {year} {2016})}\BibitemShut {NoStop}%
\bibitem [{\citenamefont {Ripoche}\ \emph {et~al.}(2017)\citenamefont
  {Ripoche}, \citenamefont {Lacroix}, \citenamefont {Gambacurta}, \citenamefont
  {Ebran},\ and\ \citenamefont {Duguet}}]{ripoche17a}%
  \BibitemOpen
  \bibfield  {author} {\bibinfo {author} {\bibfnamefont {J.}~\bibnamefont
  {Ripoche}}, \bibinfo {author} {\bibfnamefont {D.}~\bibnamefont {Lacroix}},
  \bibinfo {author} {\bibfnamefont {D.}~\bibnamefont {Gambacurta}}, \bibinfo
  {author} {\bibfnamefont {J.-P.}\ \bibnamefont {Ebran}}, \ and\ \bibinfo
  {author} {\bibfnamefont {T.}~\bibnamefont {Duguet}},\ }\href@noop {}
  {\bibfield  {journal} {\bibinfo  {journal} {Phys. Rev. C}\ }\textbf {\bibinfo
  {volume} {95}},\ \bibinfo {pages} {014326} (\bibinfo {year}
  {2017})}\BibitemShut {NoStop}%
\bibitem [{\citenamefont {Gilmore}(2008)}]{Gilmore2}%
  \BibitemOpen
  \bibfield  {author} {\bibinfo {author} {\bibfnamefont {R.}~\bibnamefont
  {Gilmore}},\ }\href@noop {} {\emph {\bibinfo {title} {Lie Groups, Physics,
  and Geometry: An Introduction for Physicists, Engineers and Chemists}}}\
  (\bibinfo  {publisher} {Cambridge University Press},\ \bibinfo {address}
  {Cambridge},\ \bibinfo {year} {2008})\BibitemShut {NoStop}%
\end{thebibliography}%

\end{document}